\documentclass[12pt]{JHEP3} 
\usepackage{epsfig}
\epsfclipon

\newcommand{\bright}{\begin{flushright}}
\newcommand{\eright}{\end{flushright}}
\newcommand{\bminip}{\begin{minipage}}
\newcommand{\eminip}{\end{minipage}}
\newcommand{\bcent}{\begin{center}}
\newcommand{\ecent}{\end{center}}

\def\ie{{\rm i.e.}}  
\def\eg{{\rm e.g.}}  
  
\def\etal{{\it et al.\/}}  

\def\be{\begin{equation}}
\def\ee{\end{equation}}
\def\bea{\begin{eqnarray}}
\def\eea{\end{eqnarray}}

\def\nn{\nonumber}
\def\ni{\noindent}

\title{Cycling in the Throat}

\author{Damien Easson$^a$, Ruth Gregory$^a$, 
Gianmassimo Tasinato$^b$~and Ivonne Zavala$^a$ \\
 ${}^a$Centre for Particle Theory, Durham University,
South Road, Durham, DH1 3LE, UK.\\ \\
${}^b$The Rudolf Peierls Centre for Theoretical Physics, Oxford University,
Oxford OX1 3NP, UK \\ \\
E-mail: \email{damien.easson@durham.ac.uk}\,, \email{r.a.w.gregory@durham.ac.uk}\,, \email{tasinato@thphys.ox.ac.uk}\,, \email{ivonne.zavala@durham.ac.uk}}

\preprint{DCPT-07/03 \\ IPPP-07/02  \\ \hepth{0701252}}

\abstract{
We analyse the dynamics of a probe D3-(anti-)brane propagating 
in a warped string compactification, making use of the Dirac-Born-Infeld 
action approximation.
We also examine the time dependent expansion  of such moving 
branes from the ``mirage cosmology'' perspective,
where cosmology is induced by the brane motion
in the background spacetime. A range of physically
interesting backgrounds are considered: AdS$_5$, Klebanov-Tseytlin and 
Klebanov-Strassler.
Our focus is on exploring what new phenomenology is obtained from
giving the brane angular momentum in the extra dimensions. We find
that in general, angular momentum creates a centrifugal barrier,
causing bouncing cosmologies. More unexpected, and more interesting, is
the existence of bound orbits, corresponding
to cyclic universes. }

\begin{document}


\section{Introduction}

One of the biggest challenges remaining for early universe cosmology is
to find a compelling explanation for inflation, or rather, an explanation
for the initial perturbation spectrum of our universe, as inferred from
the microwave background. Until recently, models for inflation were
somewhat empirically motivated, using various phenomenological scalar
fields, with parameters almost as fine tuned as those initial conditions
inflation was designed to circumvent \cite{geninfl}. 
In the past few years, string
theory has entered the fray of finding a convincing arena for inflationary
cosmology. In the past, applying string theory to the early universe was
beset by many problems, not least that of stabilising moduli \cite{modprob}.
String theory is set most naturally in more than four dimensions, and
the challenge is to explain why we do not see those extra dimensions. 
Nonetheless, applying stringy ideas to the early universe has led to many
interesting ideas, such as a possible explanation of the dimensionality
of spacetime, \cite{branvaf}, resolutions of the initial singularity
via duality \cite{stringcos}, and the idea that we do not
see the extra dimensions because we are confined to live on
a braneworld \cite{EBW,brw}.

Whether or not we live on a brane, stabilising extra dimensions
is very much an issue, and early on it was realised in supergravity that
fluxes living in the extra dimensions could stabilise the 
compactification \cite{SS}. In fact, this idea has been used to 
attempt to tune the cosmological constant in braneworld models
\cite{selftune}. Meanwhile, in string theory, the AdS/CFT
correspondence \cite{adscft} has driven an exploration of new
supergravity backgrounds, dual to various gauge theories, several
of which have the supergravity D-brane sources replaced by fluxes 
on the background manifold \cite{KT,KS}. The effect of these fluxes
is to replace the AdS horizon typically present in the supergravity
D-brane solution by a smooth throat, which, in the Klebanov-Strassler
\cite{KS} case, smoothly closes off the geometry. Given the relation,
\cite{RSads}, between the AdS/CFT correspondence and the Randall-Sundrum
\cite{RS} braneworld model, it is natural that a neat description
of hierarchies in flux compactifications in string theory should 
use the concept of warping \cite{GKP}.

Most recently, many of these ideas have dovetailed together in what is
dubbed the KKLT scenario \cite{KKLT}.
Although this description is inherently stringy, it also corresponds
to concrete supergravity realisations, and can therefore be used as
a setting for exploring possible braneworld cosmologies. For example,
the early brane inflation models \cite{brinfl} can be set in this
context to get stringy brane inflation, as in \cite{KKLMMT}.
In these scenarios, the brane location on the internal manifold 
is promoted to a scalar field in our effective cosmology, hence 
motion on the internal manifold is key to determining the 
dynamics of inflation.  Indeed, 
the warping induced by the stabilising fluxes plays a crucial role in 
providing a sufficient period of inflation: it reduces
the size of parameters  related to the inflationary potential, rendering
it  flat enough to satisfy the slow roll conditions. 

A D-brane or anti-brane wandering on a warped throat experiences 
another interesting effect, namely the existence of a ``speed limit''
\cite{st} on the motion of the brane which results from a
constraint imposed by  the non-standard form of
the  DBI brane  action. This means that the contribution 
of  the kinetic terms can become negligible compared to the
potential terms in the brane action, which then dominate
and can drive inflation.
Besides the potential cosmological applications of this effect, it is 
interesting to explore the effects of a non-standard DBI
brane action for the trajectories of branes in warped backgrounds
in more generality.

One key lesson learned from empirical braneworld models, such as
Randall-Sundrum, is that brane cosmology is achieved by having
branes moving in a curved background spacetime \cite{MOVbr}, indeed,
in the case of the RS model, the cosmological problem can be 
completely and consistently solved for the brane and bulk, as the 
system is integrable, \cite{BCG}, giving rise to a ``non-conventional''
Friedman equation on the brane \cite{NCC}. Unfortunately,
this beautiful simplicity is destroyed by the addition of
extra fields, \cite{SSG}, or extra co-dimensions \cite{HCD}, such
as would be present in a stringy compactification.
Nonetheless, the {\it mirage} approach to brane cosmology, \cite{kiritsis}
is a fascinating proposal in which we ignore the full gravitational
consistency of a particular moving brane solution (somewhat 
justified in the case of a single brane for which the supergravity
solution is somewhat suspect in any case unless at distances larger
than the compactification scale) and simply lift the concept of
cosmology as motion to higher codimension. In this picture, our Universe
is a probe brane moving in some supergravity background, and the
induced metric in the four non-compact dimensions is therefore
time dependent by virtue of this motion. This time dependent metric
therefore has the interpretation of a cosmology, and the effective
energy momentum (given by computing the Ricci curvature of this
induced metric) is the mirage matter source on the brane. This
picture -- defined and developed in \cite{kiritsis} -- has been the
basis for much work on probe-brane cosmology in string theory.

The key feature of probe brane (or indeed interacting brane)
calculations is that the location of the brane on the compact
manifold acts as a four-dimensional scalar field in our non-compact
four-dimensional Universe. This scalar field has a well
prescribed action (in the probe brane case) given by the 
Dirac-Born-Infeld action for the kinetic terms, and a Wess-Zumino
term \cite{DBI}. These together give an attractive potential
in the case of an anti-D3-brane, such as the KKLMMT
model \cite{KKLMMT}, or a velocity dependent potential in the
case of a D3-brane \cite{st}. In many ways,
this latter case is reminiscent of the ideas of kinetic 
inflation \cite{KINFL}. The idea is to use the motion of the
wandering brane on the internal manifold to generate cosmological
evolution, or (equivalently) to use the additional energy momentum
of the brane as a cosmological source, and analyse the importance
of stringy features, such as the DBI action, or particular
warped backgrounds, such as the Klebanov-Tseytlin (KT) \cite{KT}
or Klebanov-Strassler (KS) \cite{KS} throats.

The KT and KS solutions are based on the conifold
background of string theory \cite{conifold}. The 
Klebanov-Tseytlin solution describes a singular geometry produced by $N$ 
D3-branes and $M$ D5-branes wrapping a (vanishing) 2-cycle of
the conifold ({\em fractional D3-branes}). 
The Klebanov-Strassler solution is a smooth supergravity solution
in which the conifold base of KT has been replaced by the 
deformed conifold (see \cite{conifold}), giving a differential warping 
in the internal `angular' directions of the space. Both
metrics asymptote anti-de Sitter space in the UV, are
equivalent at intermediate scales, but in the IR the KT solution
has a singularity at its tip, whereas KS rounds off smoothly
(see figure \ref{KTKS})\footnote{These explicit solutions 
are of course not compact, however, they are regarded as good approximatations
in the infrared to warped throat regions of the compact Calabi-Yau (CY) 
3-manifold, and therefore reliable solutions below 
some large value of the internal radial coordinate \cite{GKP}.}.

Probe brane analysis with an eye on possible cosmological applications, 
initiated by the {\it mirage} work \cite{kiritsis,moremir}, 
was started in the context of warped compactifications
in \cite{km}, where a D3-brane moving radially
on the KS background was analysed. The key discovery was that
the brane `bounced' (\ie~simply fell to the bottom of the throat
and came back out again) and hence the cosmology passed from
a contracting to an expanding phase. This gives an alternative 
realisation to earlier superstring cosmology ideas \cite{stringcos}, 
which also featured an initial singularity avoiding bounce. Note,
this bounce is quite distinct from the ``Big Crunch -- Big Bang'' 
type of bounce typical of colliding brane scenarios \cite{ekpyrosis},
in which there is an issue, yet to be satisfactorily resolved, of
continuation through the singular collision \cite{collide}, which is
vital to obtaining the correct cosmological perturbation spectrum.

Further work on cosmological branes has attempted to take
backreaction into account \cite{malda}
(although the validity of any supergravity
approach at small scales is problematic).
For example, in \cite{st}  (see also \cite{moredbi}),
standard gravitational couplings of the brane scalar field
were considered, and candidate potential terms for the scalar field. 
In most cases, a radial motion of the brane is considered, i.e.~the 
brane simply moves `up' or `down' the warped throat. However,
since the internal manifold has five other internal dimensions, it
is natural to consider the effect of angular motion on the brane (as
originally anticipated in \cite{kiritsis}). Typically, angular
momentum gives rise to centrifugal forces, hence we might expect
brane bouncing to be ubiquitous for orbiting branes. Indeed, in 
\cite{brax} it was argued that this was the case by considering a
(Schwarzschild)
AdS$_5 \times S^5$ background. Also, in the case of {\it Branonium}
\cite{branonium}, a bound state of an orbiting anti-D3-brane, 
angular momentum was crucial in obtaining this (unstable) bound state.
(See also \cite{morebounce}.)

The aim of the present work is to extend previous analyses and study the
effects of angular momentum on the probe brane motion specifically in warped
Calabi-Yau throats. Recall that the location of
the brane on the internal manifold will become the inflaton in
a fully realistic description of brane cosmology. However, because the
angular directions in the internal manifold correspond to Killing
vectors of the geometry, the angular variables per se are not 
dynamical variables from the 4D cosmological point of view. Thus,
the inflaton in our models is precisely the same as the inflaton
in the KKLT-based models, what is new is that angular momentum
provides additional potential terms for the inflaton.
So far, angular motion has only explicitly been 
considered for the case of an AdS$_5 \times S^5$ background \cite{germani}, 
or the AdS-Schwarzschild$_5 \times S^5$ \cite{brax}, although in
\cite{germani} the slow motion (i.e.~non DBI) approximation 
in a KS throat was considered. The KS throat already exhibits 
a bouncing universe,
but we expect angular momentum to induce a bounce at a larger value
of the scale factor. We are also interested in whether there are any 
qualitatively new features for orbiting branes, for example, the
branonium is a rather different qualitative solution for an anti-D3
brane to the inflationary solution of KKLMMT.

We find a resounding affirmative outcome to these two investigations.
As with Germani \etal \ \cite{germani}, we find that the effect
of orbital motion in the DBI action is to slow down the brane radial
velocity. This slowing is crucial for having bounces as used in their
slingshot scenario. However, more interestingly we find
qualitatively new behaviour for orbiting {\it branes}, which can
now have bound states in the IR region of the KS and KT throats.
We stress that these are branes and not anti-branes, hence quite distinct
from the branonium set-up. These solutions have the cosmological interpretation
of cyclic universes (though without the big bang collision of
the more controversial cyclic scenario \cite{cyclic}).

The organisation of the paper is as follows. In the next section 
we  present the general set up we are considering. We derive the
Hamiltonian for the brane (or antibrane) moving on a generic
background. We also discuss the qualitative brane motion, and the
effect of angular momentum. We then turn to three main examples of
brane motion. In section three, we analyse brane motion in anti-de
Sitter space as a warm up exercise. In sections four and five, we then turn 
to explicit SUGRA backgrounds which modify the IR behaviour of the
AdS spacetime: the (singular) Klebanov-Tseytlin (KT), and the
(regular) Klebanov-Strassler (KS) metrics. The main result we find
is that angular momentum introduces additional turning points in the
radial motion of the brane -- allowing for bouncing, or no big bang
cosmologies. This result is no surprise, after all, angular momentum
generically introduces centrifugal barriers. However, what is
surprising is that we find regions of parameter space where bound
orbits can occur -- i.e.~cyclic cosmologies. We show that there 
are bound states of the D3 probe brane in the KS background 
which correspond to cyclic
cosmologies. In the final section we summarize and comment on
the effect of back-reaction.


\section{Probe brane analysis: General set up}

In this section we derive the general action and equations
of motion for a probe D3 or anti-D3-brane moving 
through a  type IIB supergravity 
background describing a configuration of branes and fluxes. 
A probe brane analysis is self consistent when we
consider a single brane moving in a background made by a large
numbers of other branes. Our approximations are valid provided
we remain both within the string perturbation theory regime, i.e.~the 
string coupling $g_s(r)= g_s\,e^{\phi(r)}$ is everywhere small, and
within the supergravity limit, i.e.~in regimes of
small  curvatures. In
our case, this is true provided the number of branes and/or 
fluxes in the background is large enough, so that the curvatures are
kept small. We will make this statement precise in the particular 
examples we consider in the next sections.

We start by specifying the Ansatz for the background fields we
consider, and the form of the brane action.
We are interested in  compactifications of type IIB theory,
in which the metric takes the following general form 
(in the Einstein frame)
\be\label{metrica}
ds^2= h^{-1/2}\,\eta_{\mu\nu}\,dx^\mu dx^\nu + h^{1/2}\,g_{mn}
dy^m dy^n\,.
\ee
Here $h$ is the warp factor, which in the examples we are considering,
depends only on a single, radial,
combination of the internal coordinates $y^m$, dubbed $\eta$.
The internal metric $g_{mn}$ depends on
the internal coordinates $y^m$, however, this dependence is
consistent with a compact `angular' symmetry group of the underlying
spacetime, so that particle or brane motion has conserved
angular momenta.  The field strengths  
that can be associated with a metric of this form are
$F_1$, $F_3$, $H_3$ and  $F_5$. 
These fields have only internal components, apart from
$F_5$, which can assume  the form 
$$
F_5 = dC_4 = g_s^{-1} dh^{-1}\wedge dx^0 \wedge dx^1
\wedge dx^2\wedge dx^3\ + \quad {\rm dual}.
$$
Besides these fields, the dilaton can also be active and in general
is a function of the radial coordinate only, $\phi=\phi(\eta)$. 

We now embed a probe D3-brane (or an anti-brane) in this background, 
with its four infinite
dimensions parallel to the four  large dimensions of the background 
solution. The motion of such a brane is described by the sum of the
Dirac-Born-Infeld (DBI) action and the Wess-Zumino (WZ) action. The 
DBI action is given, in the string frame, by
\be\label{dbi}
S_{DBI} = -T_3\,g_s^{-1} \int{d^4\xi\, e^{-\phi} \sqrt{-det 
(\gamma_{a b}+ {\mathcal F}_{a b} )}}\,,
\ee
where 
${\mathcal F}_{a b}={\mathcal B}_{a b} + 2\pi\alpha'\,F_{a b}$,
with ${\mathcal B_2}$ the pullback of the 2-form field to the 
brane and $F_2$ the world volume gauge field. $\gamma_{a b}
= g _{MN}\,\partial_a x^M\partial_b x^N $, is the pullback of the 
ten-dimensional metric $g_{MN}$ in the string frame. 
Finally $\alpha'=\ell_s^2$ is the 
string scale and $\xi^a$ are the brane world-volume  coordinates. 

This action is reliable for arbitrary values of the gradients 
$\partial_a x^M$, as long as these are themselves slowly varying in 
space-time, that is, for small accelerations compared to the 
string scale (alternatively, for small extrinsic curvatures
of the brane worldvolume). In addition, recall that
the string coupling at the location of the
brane should be small, i.e.~$g_s\ll 1$.

The WZ part is given by
\be\label{wz}
S_{WZ}= q\,T_3 \,\int_{\mathcal W}{C_4}\,,
\ee
where ${\mathcal W}$ is the world volume of the brane and $q= 1$ 
for a probe D3-brane and $q=-1$ for a probe anti-brane.

We are interested in exploring the effect of angular momentum
on the motion of the brane, and therefore assume that
there are no gauge fields living in the world volume of the probe 
brane, $F_{ab}=0$.  For convenience we take the static gauge, 
that is, we use the non-compact coordinates as our brane coordinates:
$\xi^a = x^{\mu=a}$. Since, in addition, we are interested
in cosmological solutions for branes, we consider the case where the
perpendicular positions of the brane, $y^m$, depend 
only on time. Thus
\be 
\gamma_{00}=g_{00} +g_{mn} \dot y^m \dot y^n\,h^{1/2}
= - h^{-1/2} \left ( 1 - h v^2 \right )
\ee 
and  ${\mathcal B}_{ab}=0$. Hence 
\be
S_{DBI}= - T_3\,g_s^{-1}\int{ d^4x \, e^{-3\phi} \sqrt{1-h\,v^2}}\,.
\ee
in the
Einstein frame\footnote{In $D$ dimensions,  to change  from the string 
frame to the Einstein frame, we have to make the transformation  
$G_{\mu\nu}^{Ein} = e^{-\lambda\phi}G_{\mu\nu}^{str}$ where $\lambda=4/(D-2)$
and we are defining the string frame such that the action scales as
$S^{str}\sim e^{-2\phi}(R+ \dots)$}.

Finally, summing the DBI and WZ actions, we have the total 
action for the probe brane 
\be
S= -T_3 g_s^{-1}\int{d^4x \, h^{-1}\left[  e^{-3\phi} \,\sqrt{1-h\,v^2}
  - q \right]}\,.
\ee
This action is valid for arbitrarily high velocities. Note,
the expression for the brane acceleration, as defined in \cite{st}, is
\be
a \sqrt{\alpha'}\,=\,h^{\frac14}\frac{d}{d t} \left( h^{\frac12}
v \right)
\ee
we recall that this has to be  small compared to the string scale  for
the configurations we are going to analyse. 
This can be simplified using (\ref{energy1}), to:
\begin{eqnarray}
a \sqrt{\alpha'}&=& h^{\frac14}\frac{d}{d t} \left( 
\frac{h \varepsilon\left(h \varepsilon+2 q\right)}{\left(h
\varepsilon+q\right)^2}
\right)^{\frac12}\,\nonumber\\
&=& \frac{2 h' h^{\frac14} \varepsilon }{(\varepsilon h+q)^2}\,\dot{\eta}
\,\nonumber\\
&=& \frac{2\, h' h^{\frac14}\varepsilon (g^{\eta\eta})^{1/2}}
{\left(\varepsilon h + q\right)^3} 
\left[ \varepsilon\left( h \varepsilon +2q \right)- \ell^2(\eta)
\right]^{1/2} \,.
\label{accbr}
\end{eqnarray}

\smallskip

\subsection{Brane Dynamics and the Effect of the Angular Momentum}\label{setup}

The  above action can be interpreted as describing the dynamics of a 
particle of  mass $m$ moving in the internal transverse space 
dimensions, with Lagrangian\footnote{From 
now on, we concentrate on backgrounds where 
the ten dimensional Einstein and string frames coincide ($\phi=$const.). 
More general cases can  be straightforwardly  included.} 
\be\label{L}
{\mathcal L} = -m\, h^{-1}\left[\sqrt{1-h\,v^2}
- q \right] \,.
\ee
Here, $m= T_3\,V_3g_s^{-1}$, and $V_3=\int{d^3x}$ is the volume
of the D3 brane. 
The non canonical form of the kinetic terms has interesting effects
for the dynamics of the brane, especially in the regime 
where the warp factor $h$ is large.  Indeed,  the quantity
$1-h v^2\ge 0$ must remain positive in order to have a real Lagrangian, 
and this imposes a bound on the brane ``speed" $v$. 
Note that while the brane speed depends on both the radial and
the angular coordinates of the manifold, it is only the radial 
coordinate $\eta$, and its velocity that is of cosmological importance, 
since that is what becomes the inflaton. Therefore, in what follows, we 
focus on the properties of the radial velocity.

The aim of the present work is to investigate some consequences of the
DBI-form of the Lagrangian when we allow the brane to move  along
the internal `angular' directions. 
Clearly, there will be conserved angular momenta:
\be\label{elem}
l_r  \equiv \frac{1}{m}\frac{\partial {\mathcal L}}{\partial
\dot y^r} =   \frac{g_{rs}\,\dot y^s}{\sqrt{1-hv^2}}  
\,,
\ee
corresponding to the conserved quantum numbers of the symmetry
group (here, the latin indices $r,s$ refer only to angular
coordinates, whereas the latin indices $m,n$ refer to all the internal
coordinates). 
In addition, the
energy (per unit mass), defined from the Hamiltonian 
of the system is conserved, and defined by
\be\label{energy}
\varepsilon \equiv \frac{E}{m} =  \rho_\eta\, {\dot\eta}
+ l_r\, \dot y^r  - \frac{{\mathcal L}}{m}\,.
\ee
where $\rho_\eta$ is the canonical momentum associated to the 
radial coordinate
\be\label{rhoeta}
\rho_\eta= \frac{g_{\eta\eta}\,\dot\eta}{\sqrt{1-hv^2}}\,.
\ee
Thus the velocity is 
\be\label{vee}
v^2= g_{\eta\eta}\dot\eta^2 + g_{rs}\dot y^r \dot y^s \,.
\ee
Using (\ref{elem}), the expression for the velocity can be rewritten 
as a function only of the coordinate $\eta$,
\be
v^{2}=\frac{g_{\eta \eta}\dot{\eta}^{2} +\ell^2(\eta)}
{1 + h\ell^2(\eta)}\,,\label{veleta}
\ee
where 
\be
\ell^2(\eta)=
g^{r s}l_r l_s\,.\label{defell}
\ee
This expression implies
that  the requirement  $1-hv^2>0$ leads
to a bound on the radial velocity, independent of the angular momenta,
\be\label{limit1}
h\,g_{\eta\eta} \dot \eta^2<1 \,.
\ee
This constraint on $\dot{\eta}$ implies that in regions 
where $h$ becomes large, the brane decelerates. 
Moreover, as we now discuss,  conservation of energy imposes a different 
constraint on $\dot{\eta}$ that depends on the
size of the  angular momenta.

Using (\ref{L}) and (\ref{elem}) the energy of the system can be rewritten as:
\be\label{energy1}
\varepsilon = \frac{1}{h}\left( \frac{1}{\sqrt{1-hv^2}} - q \right) \ge 0\,.
\ee
Notice that this quantity is always positive, the inequality being 
saturated when the velocity, $v$, is zero for branes ($q=1$), as expected.
Notice also that the energy increases as one approaches the speed limit
$h v^2=1$. 
Using (\ref{veleta}), we obtain the energy as a function of $\eta$,
${\dot\eta}$, and the angular momenta:
\be\label{energy2}
\varepsilon = \frac{1}{h}\left( \sqrt{\frac{1+ h \ell^2(\eta)}
{1-h\,g_{\eta\eta}\,{\dot\eta}^2}}  - q \right)\,.
\ee
Alternatively, for a given energy, $\varepsilon$, we can write the
radial velocity as
\be\label{etadot1}
\dot\eta^2 = \frac{g^{\eta\eta}[\varepsilon(h\,\varepsilon+2 q)\, 
-  \ell^2(\eta)]}{(h\,\varepsilon + q)^2} \,.
\ee

Equation (\ref{etadot1}) shows
that  the angular momentum has the effect of reducing
the radial velocity of the brane, since the terms
proportional to the angular momentum appear with a minus sign.
The angular momentum provides a new way to slow the radial 
motion of the brane, in addition to that identified in \cite{st}. 
Interestingly, when angular momentum is  switched on,  
there is the possibility that the radial speed of a brane (with $q=1$) vanishes
at finite values of the radial coordinate $\eta$, allowing for a bouncing
behaviour.
We analyse in specific examples how the angular momentum 
affects the radial motion of the brane in the background, in particular,
we focus on examples of brane trajectories with bounces and cycles
that are absent when the angular momentum is switched off.

Using (\ref{etadot1}), we can implicitly
get the time dependent evolution of the brane by integration:
\be
t-t_0 = \int{ d\eta \,\frac{(h\varepsilon + q)}
{\sqrt{g^{\eta\eta}[\varepsilon(h\,\varepsilon+2 q)
- \ell^2(\eta)]}}}\,.
\ee

For  general supergravity backgrounds with generic angular
momentum, we cannot solve for the brane motion analytically, 
but numerical solutions can be obtained.
We can also extract a great amount of qualitative information
from  (\ref{etadot1}), which allows us 
to identify bouncing or  cyclic brane trajectories.
A natural way to interpret the motion of the brane 
in a given background is obtained by 
writing  eq.~(\ref{etadot1}) as
\be
\label{defQ}
Q(\eta) \equiv \dot \eta^2= 
\frac{g^{\eta\eta}[\varepsilon(h\,\varepsilon+2 q)\,
- \ell^2(\eta)]}{(h\,\varepsilon + q)^2} \,.
\ee 
Thus, the
only physical regions where the brane can move with a given velocity 
$\dot\eta$, are those in which $Q\ge 0$. 
Given this (without specifying the 
background) we can argue that:

\begin{itemize}

\item  When the angular momentum vanishes, $Q$
is always positive  for a brane ($q=1$),  and the inequality saturates 
only when $\varepsilon=0$.  For an anti-brane ($q=-1$),
on the other hand, $Q\ge0$ only for  $h\varepsilon \ge 2$ 
The  point(s) where this inequality is saturated, correspond to 
{\em turning points} where the anti-brane stops and bounces.

\item  Once the angular momentum is turned on, one cannot have solutions
with zero energy $\varepsilon$.
Moreover, as we mentioned, the addition of angular momentum 
has the important effect of {\em slowing down} the radial 
velocity $\dot\eta$ of the brane with $q=1$,
allowing it to eventually stop and bounce.
If there are two zeros of $Q$ between which $Q>0$, 
the brane  oscillates between those turning  points forming a cyclic
trajectory.

\item One expects that near the bouncing points the brane experiences 
deceleration since the brane's radial speed is small near those points. 
By (\ref{accbr}), the acceleration of the brane around the
bouncing points is  naturally  small   since $\dot{\eta}$ is
small there.

\end{itemize}

It is interesting to note that we can reformulate the previous considerations
in a slightly different language. Rewriting
eq.~(\ref{etadot1})  in the following form
\be\label{etadot2}
\dot\eta^2 +  V_{eff} =0 \,,
\ee
where  $Q=-V_{eff}$ and defining
\be\label{veff}
V_{eff}\equiv V_\eta(\eta) + V_l(\eta) = 
- \frac{\varepsilon(h\,\varepsilon+2 q)\,g^{\eta\eta}}{(h\,\varepsilon + q)^2} 
+ \frac{g^{\eta\eta}\,\ell^2(\eta)}{(h\,\varepsilon + q)^2} \,,
\ee
we can interpret (\ref{etadot2}) as the equation describing 
the motion of a particle of mass $m=2$ and zero energy,  
in an effective potential given by the sum of a {\em radial}, 
$V_\eta$, and an {\em angular}, $V_l$, potential. We are
therefore  interested in the physically relevant regions 
where $V_{eff}\le0$.
Notice that the angular part, $V_{l}$, has the opposite
sign to the radial part, $V_{\eta}$. The contribution 
from the angular momentum is a `centrifugal' potential, 
which contributes with the opposite sign to the `radial' potential.  
The turning points correspond to values of $\eta$ for which $V_{eff}$ vanishes.

\subsection{Induced Expansion: a view from the
brane}\label{inducexp}

We now explore how an observer on the brane experiences
the features described so far, with respect to the motion of the brane
through the bulk. 
In this section, we provide the tools for analysing this issue on
supergravity backgrounds that satisfy our metric Ansatz.
In general, one finds that the metric on the brane assumes an FRW form,
and the evolution equation for the scale factor can be re-cast in
a form that resembles a Friedmann equation.

Indeed, it is
well known that the brane motion can be interpreted as cosmological 
expansion from a brane observer point of view~\cite{kiritsis}.
The projected metric in four dimensions  is given by 
\be
ds^2 = h^{-1/2}\left(-(1-h\,v^2)\,dt^2 + dx_i dx^i\right)
= -d\tau^2 + a^2(\tau)dx_i dx^i\,,
\ee
This metric has precisely a FRW form, with scale factor given by  
\be
a(\tau)= h^{-1/4}(\tau)\,\label{scfad}
\ee
and with the brane cosmic time  related to the bulk time coordinate 
by
\be
d\tau = h^{-1/4}\sqrt{(1-h\,v^2)}\,dt \,.
\ee
Using this information, we derive an expression for  the induced
Hubble parameter on the brane:
\be
H_{ind} = \frac{1}{a}\,\frac{d\,a}{d\tau} = \frac{1}{a}\, 
\frac{h^{1/4}}{\sqrt{(1-h\,v^2)}}\,\frac{d\eta}{dt}\,\frac{d\,a}{d\eta}
= -\frac{1}{4}\frac{h^{-3/4}}{\sqrt{(1-h\,v^2)}}\,
\frac{d h}{d\eta}\frac{d\eta}{dt}\,\,.
\ee
Using (\ref{energy1}) and (\ref{etadot1}), we can rewrite the Hubble parameter 
in terms of all the conserved quantities in a
way that resembles a Friedmann equation:
\be\label{hubble1}
H^2_{ind} \,= \,\left(\frac{h'}{4\,h^{3/4}} \right)^2 (h\varepsilon + q)^2 \, Q
\,=\,\left(\frac{h'}{4\,h^{3/4}} \right)^2 g^{\eta\eta}
\left[  \varepsilon\,(h\,\varepsilon +2q) -  \ell^2(\eta)  \right] \,,
\ee
where in the last equality we have used the quantity
$Q$ defined in (\ref{defQ}). Equation (\ref{hubble1}) suggests
that  an observer on the brane can experience a bouncing behavior.
A bounce corresponds to a point where  the Hubble parameter vanishes 
for non singular values of the scale factor. 
In particular, $H_{ind}$ can vanish:

\begin{itemize}
\item When the quantity $Q$ vanishes. Thus, the 
turning points found in the trajectories in the previous
section provide some of the bouncing points from the 
brane observer's point of view.

\item When $h'$ vanishes. When the metric function has an 
extremum it provides an additional possible bouncing point for an 
observer on the brane. Such points, do not, in general, 
correspond to  turning  points for 
the brane trajectories from a bulk point of view, but rather,
the motion of the brane through a turning point of $h$ causes the
induced scale factor to experience a bounce.

\end{itemize}

It is useful to point out that, although  an observer on the brane does 
experience bouncing or cyclic expansions, it does so in a frame
that is not the Einstein frame \cite{km}. Indeed, due to the presence
of the warp factor $h^{-\frac12}$ in front of the four dimensional metric, the 
four dimensional Planck
scale depends on the position of the brane in the bulk. Nevertheless,
the frame we are working in has the important feature of being
the frame in which the masses of the fields
confined on the brane are constant, in the sense that they do not depend
on the brane position.

In the next sections,  we  consider some concrete examples 
where we apply the general analysis presented in this section. 
In several situations it is possible to get bouncing 
as well as cyclic universes for an observer confined on the brane. 


\section{AdS Throat: a simple example}

As a warm up, and to show how our general discussion in the previous
section can be applied,
we start by considering, the simple case of an 
AdS$_5\times S^5$ background, 
corresponding to the near horizon limit of a stack of 
$N$ D3-branes. This case has been considered also 
in \cite{kiritsis}, \cite{germani}.
The 10D metric takes the form
\be
ds^2 = h^{-1/2} \,dx_\mu \,dx^\mu  + h^{1/2}\,(d\eta^2 + \eta^2\,d\Omega^2_5)
\ee
where $$h = \frac{\lambda}{\eta^4}$$ 
with $\lambda= 4\,\pi\,\alpha'^2 g_s N$;
$d\Omega^2_5$ corresponds to the metric of a round $S^5$ sphere. 
Studying the brane trajectory through this background, we find that
thanks to the angular momentum, the brane 
can experience a bounce,  at a location  depending
on $\lambda$, the brane energy and angular momentum.

\subsection{Brane Evolution and Physical Consequences}

Consider a probe brane (or antibrane)  moving along one of the
angular coordinates of the sphere $S^5$, $\theta$, say, and the radial
coordinate $\eta$. Thus the velocity takes the form:
\be
v^2 = \dot \eta^2 + \eta^2 \,\dot \theta^2\,.
\ee
Using the general equations in 
section \ref{setup},  the radial velocity  (\ref{etadot2}) becomes
\be
Q=\dot\eta^2 =  \frac{\varepsilon(h\,\varepsilon+2 q) }{(h\,
\varepsilon + q)^2} 
- \frac{l^2}{\eta^2\,(h\,\varepsilon + q)^2} \label{etadotADS}
\ee
with $h=\lambda/\eta^4$. 
This expression can be formally integrated in terms of elliptic 
functions, however, its form is not particularly illuminating and 
general features are easily extracted.
The bouncing points in which the
quantity $Q$ vanishes are easily found:
\be\label{roots}
\bar{\eta}^2_{\pm}=\frac{l^2 \pm \sqrt{l^4 -  
8\,\varepsilon^3\,\lambda\,q }}{4 \varepsilon\,q} \,.
\ee
The number of real positive roots depends on the angular momentum, and 
is different for a brane $q=1$ and an antibrane $q=-1$, as we now explain.

\bigskip
\ni
{\it Brane Dynamics ($q=1$)}
\bigskip

\noindent
In the case of a brane, the number of real, 
positive roots of (\ref{etadotADS}),
depends on the value of the angular momentum. 
Therefore, there are three different 
possibilities for the brane trajectories, which depend on the 
number of zeros of $Q$. Such points correspond to the turning points
discussed in sec.~\ref{setup}. For 
$l <l_c \equiv(8\,\varepsilon^3\lambda)^{1/4} $, there are no real solutions 
to (\ref{roots}),
thus the radial brane speed is always positive and the brane can move in the 
whole AdS geometry. For $l =l_c  $, there is a single 
repeated root of (\ref{etadotADS}) at $\bar\eta$, and hence 
a zero of ${\dot\eta}$ and ${\ddot\eta}$.
Thus a brane travelling toward the horizon from the UV region will actually
decay into a `circular' orbit at $\eta = {\bar \eta}$. This type of orbital
motion is counter-intuitive from the point of view of standard particle
motion, however, it is simply a consequence of the fact that our kinetic
action is DBI, and not the conventional $\frac{1}{2} v^2$.
Finally, for $l >l_c  $, there are two positive real 
roots, $\bar\eta_\pm$, that is, two turning points. 
In this case, a brane coming from infinity toward the horizon
is not able to approach closer than $\bar\eta_+$, where it
rebounds back to the UV region.  There is also an internal region, 
$\eta<\bar\eta_-$, where a brane can be bounded to travel a maximal distance 
$\bar\eta_-$ away from the horizon, where it bounces back. 

In the vicinity of the turning points, we can find analytic solutions for 
equation (\ref{etadotADS}),  by writing
\be 
\eta(t) \equiv \bar{\eta}_+ +  \delta\eta(t)
\ee
and looking for  solutions at small $\delta\eta(t)$. 
Expanding (\ref{etadotADS}) for small $\delta \eta$ yields:
\be
\delta \dot{\eta}^2 \,=\, \frac{4\bar{\eta}_+^5 \delta \eta}
{\left(\lambda \varepsilon +\bar{\eta}_+^4\right)^2}\,\left[
\left ( 4 \varepsilon \bar{\eta}_+^2-l^2 \right)
+2\varepsilon \bar{\eta}_+ \delta\eta \right ]\,.
\label{perteqADS}\ee
When $\left( 4 \,\varepsilon \,\bar{\eta}_+^2-l^2
\right) \neq 0$, we take only the dominant
first term inside the squared root in  
(\ref{perteqADS}). The solution of the equation is 
\be
\delta \eta(t) = \left(\frac{\bar{\eta}_+^5}{\left(\lambda \varepsilon
+\bar{\eta}_+^4\right)^2}\right)^2\,
\left(
4 \varepsilon \bar{\eta}_+^2-l^2
\right)^2\,t^2 \,.
\ee
from which it is apparent that the solution $\eta(t)$ has
a  bouncing behaviour for small $t$. For
$\left( 4\, \varepsilon \,\bar{\eta}_+^2-l^2 \right) = 0$, we see
the orbital relaxation to $\eta = {\bar\eta}$.

\FIGURE[ht]{\epsfig{file=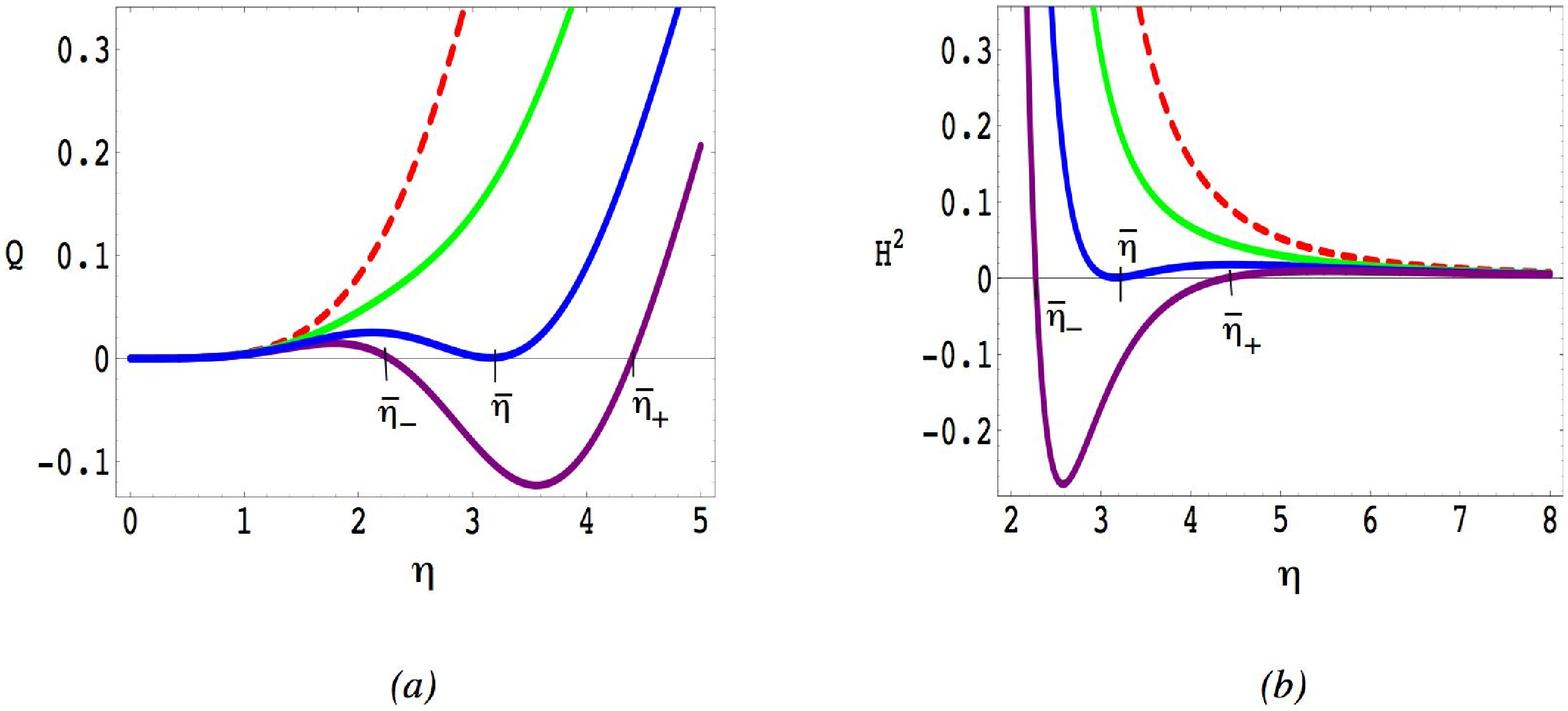, width=.9\textwidth}
\caption{Numerical analysis of the radial velocity, $Q$, and  
the Hubble induced parameter  $H^2$, for a brane wandering  in
an AdS throat  geometry.  We took $\alpha'=1$, $g_s\,N\sim \mathcal
O(10) $, and $\varepsilon=1$.  
In $(a)$ we show the evolution of $Q$ as we change the value 
of the angular momentum. The 
dashed curve corresponds to zero angular momentum. 
The blue line corresponds to the 
critical value of $l_c\sim 6.32$, for which there is only one root at 
$\bar\eta$. When the angular momentum is increased beyond this value
this root splits  into $\bar\eta_\pm$, as shown in the plot. In $(b)$
we plot the Hubble parameter  
induced on the brane.}
\label{veffplot1}} 

In figure~\ref{veffplot1}$a$ we plot the kinetic function, $Q$, for different 
values of the angular momentum for the case of a brane ($q=1$). When 
we increase the value of the angular momentum, one or 
two {\em turning points} arise,
located at the positions given in (\ref{roots}).
The value $\eta=0$, which corresponds to 
the horizon of AdS, {\em is not} a zero of $Q$ (although it is not clear 
from the figure!).  However, when the angular momentum gets  
large enough, the second 
term in the squared root in (\ref{roots}) is negligible 
and therefore the origin 
becomes eventually a zero of the potential, but the brane 
is confined to move in 
the region $\eta^2>\bar\eta^2\sim2\,l^2$. 
\FIGURE[ht]{\epsfig{file=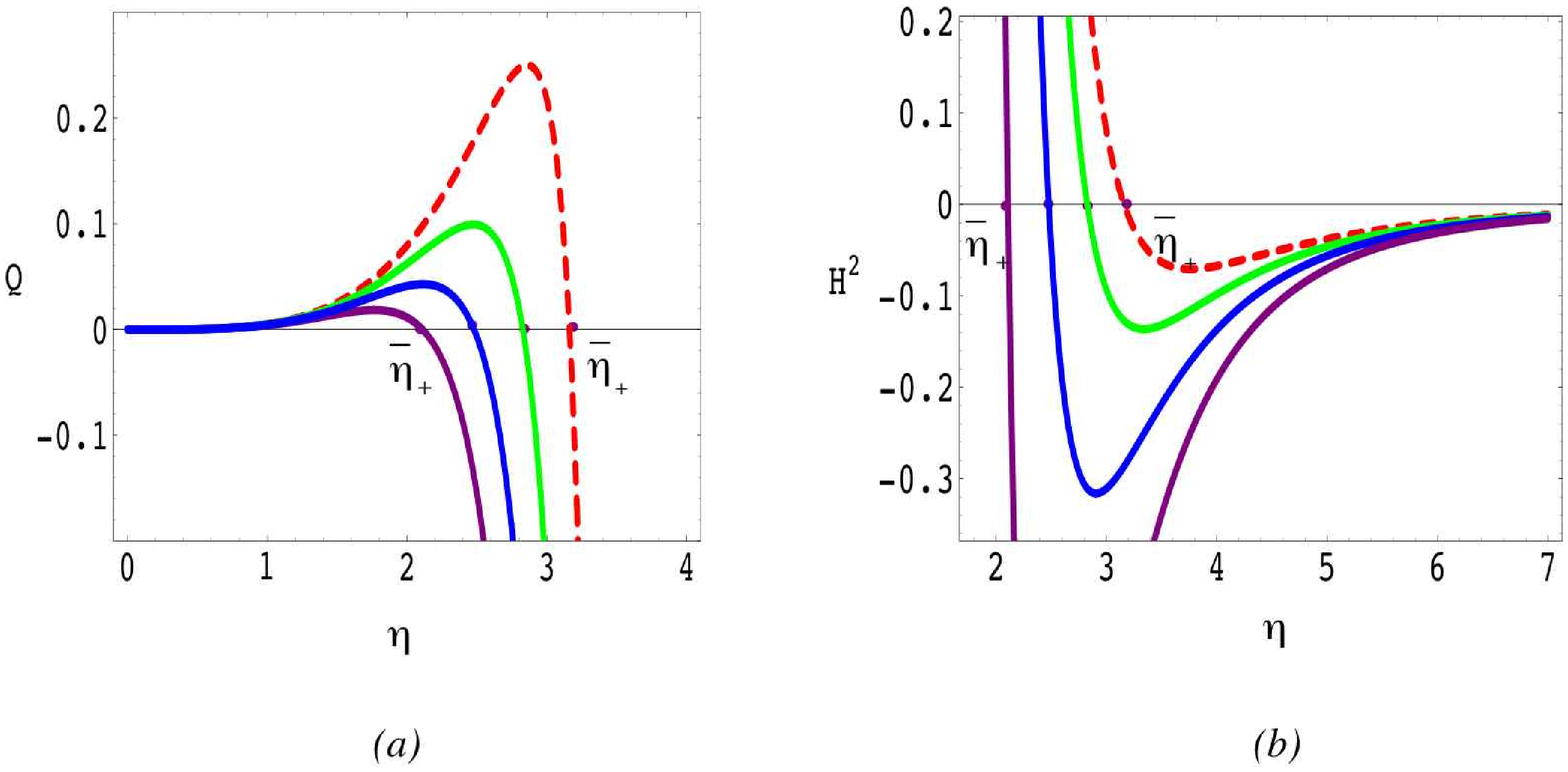, width=.9\textwidth}
\caption{Numerical analysis of $Q$ and $H^2$  for a wandering
anti-brane in an AdS throat for different values of the angular momentum. 
We take the same values as in the previous plot for the parameters. 
In $(a)$ we plot the value of $Q$ for increasing values of the angular
momentum.  The dashed line corresponds to  zero angular momentum.
In picture $(b)$ the corresponding Hubble 
parameter is shown.}\label{veffplot2}}

\bigskip
\ni
{\it Anti-Brane Dynamics  ($q=-1$)}
\bigskip

\noindent
In the case of an antibrane, there is only one real, positive solution to 
(\ref{roots}).
Thus, there is always a turning point and in fact, the antibrane can only 
{\em live} in the region  $\eta \le\bar \eta_+$. This region gets reduced as 
we increase  the angular momentum,
until it eventually disappears completely for very large angular momentum. 
In figure~\ref{veffplot2}$a$ we show an explicit numerical example, 
for the same values of the 
parameters as in figure~\ref{veffplot1}, for the case of 
an antibrane ($q=-1$). 

\subsection{Induced Expansion: Bouncing Universe}

The induced expansion in the present simple background, is predicted from 
our general discussion in sections~\ref{setup} and \ref{inducexp}. The 
form of the induced Friedmann equation in this case is \cite{kiritsis}
\be\label{hubbleads}
H^2_{ind} \,= \,\frac{\lambda^{1/2}}{\eta^4} 
\left[  \frac{\varepsilon^2\lambda}{\eta^4}\, + 2\varepsilon \,q -  
\frac{l^2}{\eta^2}\right]\,.
\ee
The induced expansion that a brane (antibrane) observer experiences,
can be summarised  as follows.

\bigskip
\noindent
 {\it Moving Brane}
\bigskip 

\noindent
We have seen that $Q$ has 
no zeros when $l=0$. Since $h'=0$ has no solutions, we conclude 
that for zero angular momentum it is not possible to have a bounce. 
However, as the angular momentum is turned on, one or two zeros of $Q$ can 
arise, that is, there can be up to two turning points, $\bar\eta_\pm$. Since 
$H^2_{ind} >0$ at infinity and for small values of $\eta$, one concludes 
that a brane arriving from large values of $\eta$ will encounter a 
turning point before reaching the AdS horizon, 
and will bounce back \cite{kiritsis,germani}. 
Since there can be two zeros, there also exists an internal 
region where a brane 
coming from the horizon will reach a given point and bounce back into 
the horizon.
This behaviour is shown in figure \ref{veffplot1}$b$.

\bigskip
\noindent
{\it Moving Anti-Brane}
\bigskip

\noindent
The case of the anti-brane is slightly 
different. There is always one zero of $Q$, with or without 
angular momentum. 
Given that, $H^2_{ind} <0$ at infinity, the only possible solution is the
internal bounce where the antibrane goes back to the horizon. 
This behaviour is shown in figure \ref{veffplot2}$b$.

\bigskip

Therefore, we conclude that in this simple case, the effect of the angular 
momentum is important to allow for a bouncing brane universe. However, a 
cyclic universe is clearly not possible. We will see in the next two sections 
that as soon as one has  more complex backgrounds, the angular momentum gives 
the new possibility of cyclic universes.


\section{Klebanov-Tseytlin Background}

The first example in which the brane trajectory exhibits a cyclic 
behaviour is the Klebanov-Tseytlin background. This is a supersymmetric,
singular solution representing the background associated with $N$ D3-branes
and $M$ D5-branes wrapping a  two cycle. The addition of the wrapped D5 
branes modifies the warp factor with a logarithmic correction that plays
a crucial role in our search for cyclic trajectories. Apart from cyclic
cosmologies, we find the important feature that large enough angular momenta
prevents the brane from falling into the singularity of the geometry. Although
the  brane can
move toward the singularity, it  bounces back  before reaching it returning
to  the regular region  of the geometry.

\subsection{The solution}

The Klebanov-Tseytlin \cite{KT} (KT)  flux background 
describes a singular geometry produced by a number, $N$,
D3-branes sourcing the self dual RR five form field strength $\tilde F_5$, 
and a number, $M$, D5-branes wrapping a (vanishing) 2-cycle (these are 
called {\em fractional D3-branes}). These branes source the RR three form
field strength $F_3$ and there is also a  nontrivial NSNS three form field
$H_3$. 
The RR $C_0$ form and the dilaton field $\phi$ vanish for this solution.
In order for our probe brane analysis to be valid, 
we need to have a large number of D3-branes and D5-branes. 
Also, in order for the supergravity description to be valid,
we need to have small curvatures. These two things can be realised
if we keep $g_sN\gg1$, and work in  the string perturbation
limit $g_s<1$. 

The complete  solution takes the following form \cite{KT,coneuse}
\be\label{KT1}
ds^2_{10} = h^{-1/2}(\eta)\,dx_\mu dx^\mu +h^{1/2}(\eta)
(d\eta^2 + \eta^2 ds^2_{T^{1,1}})\,.
\ee
Here $\eta$ is the  radial coordinate of the internal six dimensional
manifold, which is given by the {\em conifold}. This is a
six-dimensional cone with base $T^{1,1}$, where 
$T^{1,1}$  is an Einstein space whose metric is 
\be\label{t11}
ds^2_{T^{1,1}} = \frac{1}{9}\, (g^5)^2 + \frac{1}{6} \,\sum_{i=1}^4(g^i)^2\,.
\ee
Topologically, this is $S^2\times S^3$, and the one-form 
basis $\{g^i\}$ above is the one conventionally used in the literature,
given in terms of the angular internal coordinates as:
\bea\label{basis1}
g^{1,3} = \frac{e^1\mp e^3}{\sqrt{2}} \,\,, \qquad  
g^{2,4} = \frac{e^2\mp e^4}{\sqrt{2}} 
\,\,, \qquad g^5 = e^5\,,
\eea
where,
\bea\label{basis2}
&& e^1 = -\sin{\theta_1}\,d\phi_1\,\,, \qquad e^2 = d\theta_1 \,\,, \qquad  
e^3 = \cos{\psi}\,\sin{\theta_2}\,d\phi_2 - \sin{\psi}\,d\theta_2 \,,  \nn \\
&& e^4 = \sin{\psi}\,\sin{\theta_2}\,d\phi_2 + \cos{\psi}\,d\theta_2 \,\,,  
\qquad  e^5 = d\psi + \cos{\theta_1}\,d\phi_1 + \cos{\theta_2}\,d\phi_2 \,,  
\eea
with $0\le \psi \le 4\pi$, $0\le \theta_i \le \pi$,
$0\le \phi_i \le 2\pi$. The $S^2 \times S^3$ topology can now be readily
identified as \cite{coneuse}
$$ 
S^2: \quad \psi=0\,\,, \quad \theta_1=\theta_2\,,  
\quad \phi_1=-\phi_2\,\,;\qquad \quad  {\rm and } 
\qquad \qquad  S^3\,\,: \quad \theta_2\,=\,
\phi_2=0\,. 
$$
The other background fields are given by \cite{KT}
\bea
&& B_2 = \frac{3\,g_s\,M\,\alpha'}{4}\,\left[ 
\ln{\frac{\eta}{\tilde{\eta}}}\right]
\, (g^1\wedge g^2
+ g^3\wedge g^4)  \nn \\  
&& H_3 = dB_2 =  \frac{3\,g_s\,M\,\alpha'}{4\,\eta}\, d\eta\wedge
(g^1\wedge g^2 + g^3\wedge g^4) \nn \\
&& F_3 = \frac{M\,\alpha'}{4}\,g^5\wedge(g^1\wedge g^2
+ g^3\wedge g^4)\nn \\
&& \tilde F_5 = {\mathcal F}_5  +  \star{\mathcal F}_5 \nn \\
&& {\mathcal F}_5 = B_2\wedge F_3 = 
27\,\pi\,\alpha'^2N_{eff}(\eta)Vol(T^{1,1})  \nn \\
&&  \star{\mathcal F}_5 = dC_4 =  g_s^{-1}
d(h^{-1})\wedge dx^0\wedge dx^1\wedge dx^2\wedge dx^3  \nn
\eea
where the volume of $T^{1,1}$ is computed using the
metric (\ref{t11}) and is given by $Vol(T^{1,1})= 16\,\pi^3/27$.
The functions appearing in the solution are 
\bea
&& h(\eta) = \frac{27\,\pi\,\alpha'^2}{4\,\eta^4}\,\left[ g_s\, N +
\frac{3\,(g_s\,M)^2 }{2\pi} \,\left( \ln{\frac{\eta}{\tilde\eta}} +
\frac{1}{4}\right) \right]
= \frac{c}{\eta^4}\,(1+b\ln{\eta})\,, \label{hkt} \\
&&  N_{eff} = N + \frac{3\,(g_s M)^2 }{2\pi} \,\ln{\frac{\eta}{\tilde{\eta}}}
\,.
\eea
Here, $\eta=\tilde\eta$ determines the UV scale at which the 
KT throat joins to the Calabi-Yau space.  This solution has a 
naked singularity at the point where $h(\eta_0)=0$, located 
at $\eta_0={\tilde \eta}e^{-1/b}$.  
In this configuration,  the supergravity 
approximation is valid when $g_sM,\,g_sN \gg1$: in this limit
the curvatures are small, and we keep $g_s<1$. (By taking the 
parameter $M=0$, one finds AdS space, without singularities.)


\subsection{Brane  Evolution and Physical Consequences}

We are now ready to study the evolution of our probe (anti-) brane in this 
background. 
In the present solution, the Lagrangian describing this motion is given by 
(\ref{L})~\footnote{Note that since the dilaton is zero for this 
solution, then the Einstein and string frame coincide. }:
\be\label{Lkt}
{\mathcal L} = -m\, h^{-1}\,\left[ \,\sqrt{1-h\,v^2}
- q \right] \,.
\ee
The velocity of the brane in the 
internal space is given by 
\be\label{vel1}
\dot \eta^2  + \eta^2\left[
\frac{1}{9}(\dot g^5)^2    + \frac{1}{6}\sum_{i=1}^4(\dot g^i)^2 \right]\,,
\ee
where, with a slight
abuse of notation, that should  not generate confusion,
we have denoted
\bea
\dot g^5 = \dot \psi+ \cos{\theta_1}\dot \phi_1 + \cos{\theta_2}\dot \phi_2
\nn 
\eea
and similarly for the other $\dot{g}^{i}$.
In general,  studying the motion along all the internal coordinates  is not
simple, however, without loss of generality,
we can concentrate on the case where the brane moves
along one of the cycles of the internal manifold. In particular, let us
consider motion on the $S^3$ cycle, defined by  $\theta_2=\phi_2=0$ 
and take $\phi_1=$constant for simplicity. In this
case (writing $\theta_1 \equiv \theta$) we obtain:
\be\label{vel2}
v^2 = \dot \eta^2  + \eta^2\left[
\frac{\dot \psi^2  }{9}  + \frac{\dot \theta^2}{6} \right]\,.
\ee
Plugging this into the Lagrangian, we see that  there are two conserved
angular momenta along $\psi$ and $\theta$. 
These are 
\bea
l_{\psi} = \frac{\eta^2\dot \psi}{9\,\sqrt{1-h \,v^2}} \qquad ; \qquad
l_{\theta} = \frac{\eta^2 \dot \theta}{6\sqrt{1-h \,v^2}}  \,.
\eea
Defining the angular momentum, $l$, via
\be
l^2\equiv 9 \, 
l_{\psi}^2+ 6 \,l_{\theta}^2 \,, 
\ee
gives the velocity:
\be
v^2\,=\,
\frac{\eta^2 \dot{\eta}^2+l^2}{\eta^2+ h\,l^2}\,.
\label{velokt}
\ee
It is clear that when $l=0$ we obtain the simple expression 
$v^2=\dot \eta^2$ and the speed limit in this case is simply $h\dot\eta^2<1$.
The canonical momentum associated to $\eta$, (\ref{rhoeta}), for KT is 
given by
\be
\rho_\eta = \frac{\dot \eta}{\sqrt{1-h \,v^2}} \,. \label{rhoetaKT}
\ee 
Using this information  the energy  (\ref{energy2}) takes the form,
\be\label{energyKT}
\varepsilon = \frac{1}{h} \left[ \sqrt{\frac{1+ h\,\frac{l^2}{\eta^2}}{
1-h\, \dot\eta^2}} - q \right]\,,
\ee
that is, an expression formally identical to the one for the motion in
the $AdS_5 \times S_5$ background analysed in the previous section, 
however, note $h$ is different.  Consequently, in terms of $h$,
the expression for the time derivative of $\eta$, (\ref{etadot1}), in
the present case takes the familiar  form:
\be \label{exvelKT}
\dot \eta^2 = \frac{\varepsilon(\varepsilon h+2q) - l^2/
\eta^2 }{(h\varepsilon +q)^2}  = Q \,\,.
\ee 
We are interested in those regions of the space where
$Q\ge0$, therefore, we can simply look at the behaviour of 
\be\label{Q1}
\varepsilon(\varepsilon\,h+2q) - \frac{l^2}{\eta^2} \ge 0 \,.
\ee
Since $h(\eta)$ is more complicated than before, this is no longer 
as amenable to analytic analysis and we must proceed numerically. 
Of particular interest are those points where $Q=0$, where we have
{\em turning points} for the brane (anti-brane). 
At such values, the brane typically stops and bounces.
The number of zeros of  (\ref{exvelKT}) is obtained by  
looking at the solutions to 
\be
\varepsilon^2 h -\frac{l^2}{\eta^2} = -2\varepsilon\,q
\,,
\ee
which are easily obtained by looking
at its asymptotic behaviour, and the form of the function $h$. 

In what follows, we illustrate trajectories for a brane or 
anti-brane as it moves in a specific KT background, which is
representative of the types of behaviour one can obtain.
Figure \ref{hKT} shows the warp factor $h(\eta)$ for the
representative values of the parameters: 
$g_sN,\,g_sM\sim {\mathcal O}(10)$, $\alpha'=1$. It is important to 
notice that this function has a zero (the location of the singularity)
at the position $\eta=\eta_0$.  Moreover, $h$ has a maximum 
at the point $\eta'>\eta_0$, where  the derivative $dh/d\eta = h'=0$.
This fact will become important in the next subsection, when we
discuss the induced time dependent  expansion on the brane. 

\FIGURE[ht]
{\epsfig{file=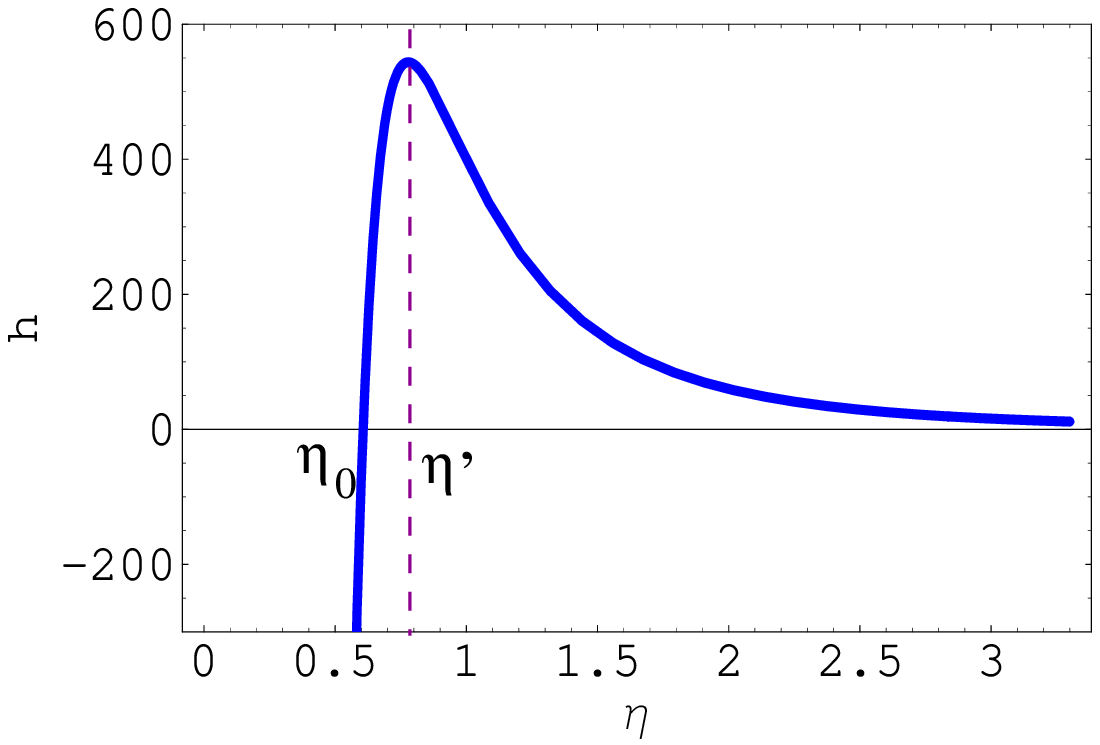, width=3.5in}
\caption{Plot of $h(\eta)$ given by Eq.~(\ref{hkt}). 
In the plot we take $\alpha^\prime=1$ and $g_s M$, $g_s N \sim
\mathcal O(10)$.  The point $\eta=\eta_0$ 
corresponds to the singularity, where $h$ vanishes. 
The maximum of $h$, is located at the point $\eta'$, defined by
$h'(\eta')=0$.}  \label{hKT}}

\FIGURE[ht]{\epsfig{file=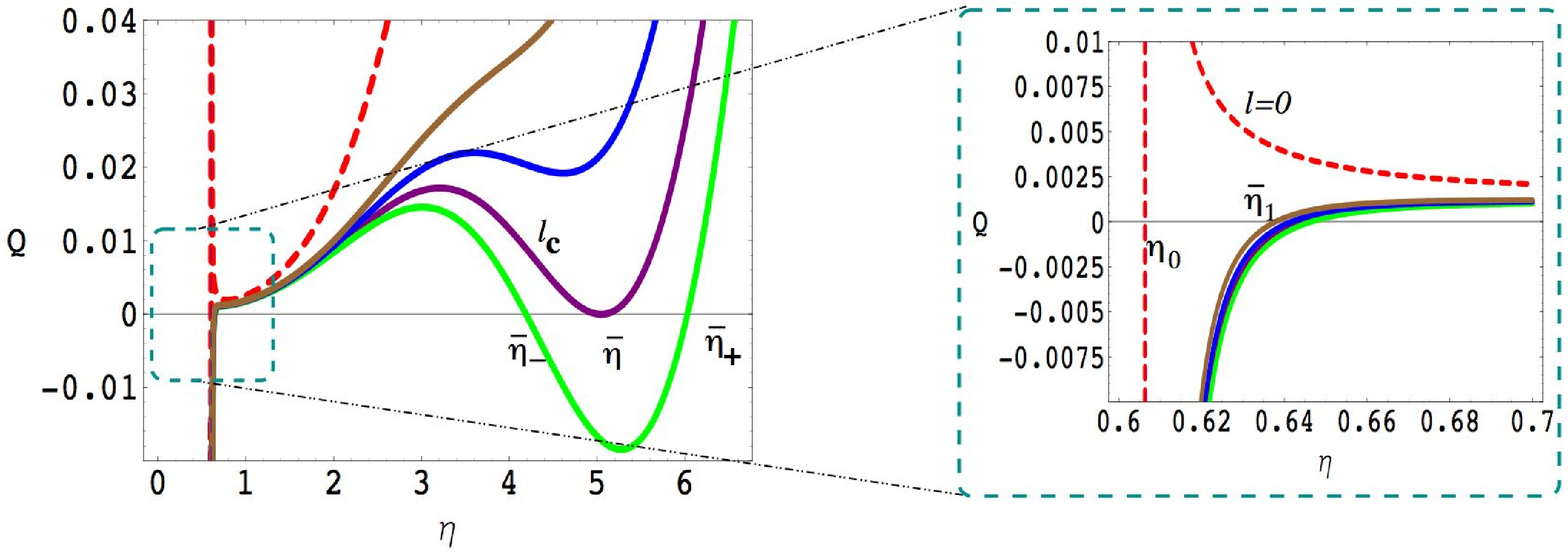, width=.9\textwidth}
\caption{$Q(\eta)$ in the KT background for various angular 
momenta. We take 
$g_s M$, $g_s N \sim \mathcal O(10)$, $\alpha^\prime =\varepsilon =
q=1$. The dashed line represents the value of $Q$ for zero 
angular momentum. The value of $l$ where a second zero of $Q$
first appears, for the choice of parameters given, is  
$l_c\sim 10.84$. The position of the first zero of $Q$, can 
be seen by zooming into the plot for small values of $\eta$. 
>From here, it is clear that $\eta_0<\bar\eta_1$, 
where $\eta_0$ is the position of the singularity, 
$h(\eta_0)=0$. 
}\label{QKT1}}

\FIGURE[ht]
{\epsfig{file=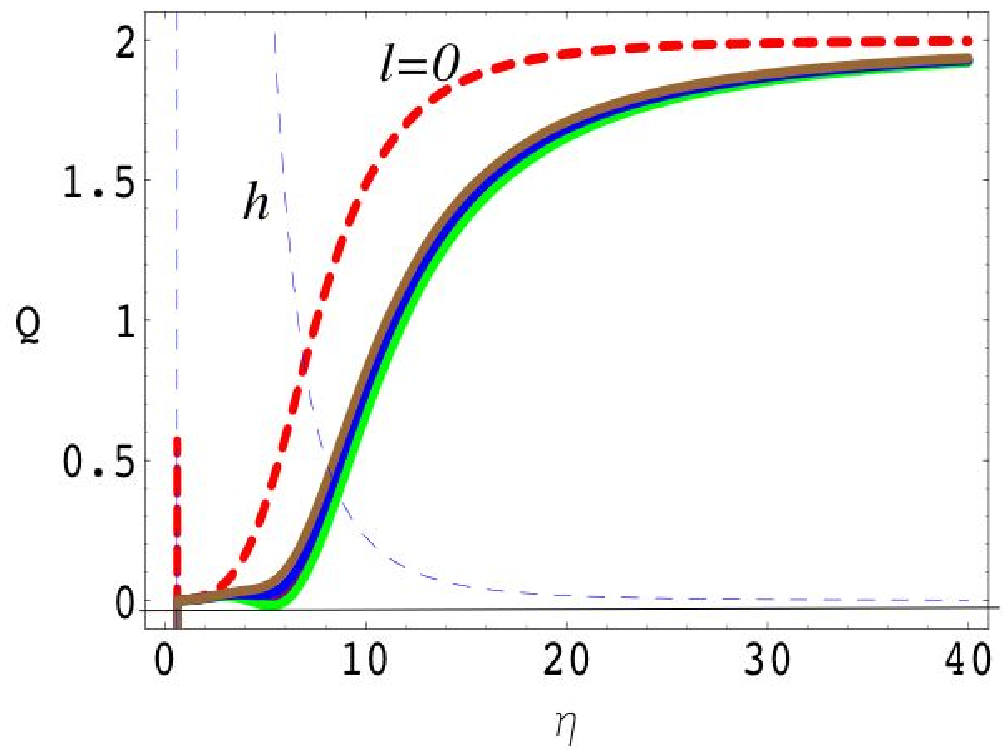, width=3.5in} 
\caption{
Large $\eta$ behaviour of $Q$ in the KT background for the same 
values of the parameters as in fig.~(\ref{QKT1}). The thin blue line
corresponds to the metric function $h$, as indicated in the figure. 
}\label{ktlarge}}

\bigskip
\ni
{\it Dynamics of a Brane ($q=1$)}
\bigskip

\ni
We start our discussion with the case of a brane ($q=1$) 
moving in the KT background.  In figure \ref{QKT1}, we show the 
behaviour of $Q$ for a probe brane
with  fixed energy, $\varepsilon=1$,  for the same
values of $M$, $N$ as in figure (\ref{hKT}). 
We show explicitly how $Q$, and consequently the brane trajectories, vary
as we change the angular momentum. The general features of the motion
of a brane with angular momentum in the KT throat are summarised as follows:

\begin{itemize}

\item  $l=0$.
For vanishing angular momentum, (\ref{exvelKT}) becomes
\be
\frac{h+2 }{(h + 1)^2}  = Q\,,
\ee 
(recall that we are taking $\varepsilon=1$). 
Therefore, at the singularity $\eta_0$ ($h=0$), 
and also at large $\eta$ ($h \rightarrow 0$) we have that $Q= 2$.
For intermediate $\eta$, the radial speed
decreases to a minimum value, defined
by $Q'=0$, where $Q'=dQ/d\eta=-h'(h+3)/(h+1)^3$. 
There are no turning points of brane motion in this case, \ie~solutions to
$Q=0$. Physically, this implies that a brane
coming from the far UV region will not be able to escape from falling
into the singularity of the KT geometry at some finite time, 
although its velocity will damp to a minimum at $\eta'$. As we now
discuss, this can be avoided by turning on angular momentum. 

\item  $l\ne 0$.  
As the angular momentum is turned on, several qualitatively new
phenomena can occur. First of all notice that 
\be
Q = \frac{h+2 -l^2/\eta^2}{(h+1)^2}\,,
\ee 
therefore at the singularity 
$Q(\eta_0) = 2 -l^2/\eta_0^2$. It is clear that as soon as we take,
$l^2>l^2_{min}=2\,\eta_0^2$,  \,\, $ Q(\eta_0) <0$ at the singularity, 
hence the singularity is now in a kinetically forbidden region. 
On the other hand, since $Q \to 2$ for large values of $\eta$,
independent of the angular momentum (see fig.~\ref{ktlarge}), we
conclude that $ Q$ has to cross zero {\em at least once} 
between the singularity and infinity. 
In other words, for generic values of the angular momentum, there is
at least one point, $\bar\eta_1>\eta_0$, where the radial brane speed 
becomes zero, $Q=0$. Therefore, a brane coming from the UV region will 
bounce back {\em before} reaching the singularity. 

The points where the radial brane speed vanishes are 
determined by the equation 
\be\label{eqgeta}
G(\eta) = 2{\eta^4}- {l^2}{\eta^2} + c(1+b\,\ln{\eta}) =  0 \,.
\ee
This equation can have a maximum of
three (real positive) zeros. The position and number of such zeros
depends on the value of the angular 
momentum\footnote{ To see this, note $G(\eta_0) < 0$ for
$l>l_{min}$ and $G \rightarrow \infty$ as $\eta \rightarrow \infty$, 
thus $G$ has at least one zero.  
Then, noting that the turning points of $G$, $G' =0$, 
are determined by a quadratic in $\eta^2$, we
see that $G$ can have at most two turning points, 
hence at most three zeros. Whether or not $G$ has these additional zeros
depends on the relative magnitudes of $l$, $b$, and $c$. We find that
for physically relevant values of $b$ and $c$, when $l$ becomes large enough
these extra zeros typically appear.}.

We can thus have three different situations, depending on the
value of the angular momentum. 
Let $l_c$ be the value of the angular momentum for
which an additional zero of (\ref{eqgeta}) appears. For
$l<l_c$, there is only  one zero\footnote{We will always consider
values of $l>l_{min} \equiv\sqrt{2}\,\eta_0$, so that $\bar\eta_1>\eta_0$. } 
at $\bar\eta_1$, say. For $l>l_c$, the
second zero $\bar\eta$ resolves into two different zeros,
$\bar\eta_\pm$ say, and we end up with three zeros. The presence of these
turning points is quite interesting as it allows for a variety of new
trajectories that the brane can follow. Also, note
that the singularity of KT is avoided for all these
trajectories.  These behaviours are important for the induced expansion,
allowing for novel bouncing and cyclic universes. 
We now turn to a discussion of the
brane physics for these values of the angular momentum. 
 
\begin{itemize}
 
\item $l_{min}<l<l_c$. 
In this case, a D3-brane falling into the KT throat decelerates
more rapidly than if it had no angular momentum, and eventually 
slingshots at $\eta=\bar\eta_1$, rebounding back  up
the throat, {\em without} ever reaching the singularity.

\item $l=l_c$. In this case, a new zero for the
radial speed arises at $\bar\eta=\bar \eta_+ = \bar \eta_-$. 
Therefore, a brane coming down
the throat asymptotes a steady orbit at that point. 
Moreover, a new kind of trajectory is  possible. 
A brane can actually be in orbit (at $\bar \eta$) around the 
KT singularity and fall in towards the conifold tip, rebound 
and asymptote its original orbit.

\item  $l>l_c$.  For  larger values of the angular momentum,
there are two {\em separate} physical regions for the 
brane motion. 
For the region $\eta>\bar\eta_+$, the brane descends 
through the throat, from large to small values of $\eta$, 
slingshotting at $\bar{\eta}_{+}$, and rebounding back toward 
large values of $\eta$. 
On the other hand, in the region $\bar\eta_1< \eta <\bar\eta_-$, 
a {\em bound state} arises, corresponding to the brane 
oscillating between these two turning points (see fig.~\ref{QKT1}). 
The presence of these 
turning points will be very important for the induced expansion 
analysis in the next section, as it gives rise to cyclic universes.
 
\end{itemize}
\end{itemize}

\bigskip
\ni
{\it Dynamics of Anti-Branes ($q=-1$)}
\bigskip

Let us now move on to the case of an anti-brane moving along the KT throat. 
This  is, in some way, simpler than the brane case, as we now see. 
We show this example numerically in figure~\ref{QhubbleKTq-1}$a$, 
for the same values for the parameters as in figures
\ref{hKT}--\ref{ktlarge}. 

\begin{itemize}
 
\item $l=0$. When no angular momentum is present,  
$Q=  (h-2)/(h-1)^2$. 
Therefore, this quantity has a singularity at the two points where $h=1$. 
When $1<h<2$, $Q<0$, and therefore this region is kinetically forbidden. 
When $h>2$, $Q$ becomes positive, however, for large values of $\eta$,
it becomes negative (unphysical) again, as $Q\to -2$. It is clear that 
$Q=0$ has two zeros, corresponding to turning points of the motion,  at the
positions where $h=2$.  The precise location of these  
is  determined by the equation,
\be
\frac{c}{\eta^4}\,(1+b\,\ln{\eta}) =2 
\ee
and we call $\bar\eta_\pm$ the positions of these zeros. 
Of course, the regions beyond the greatest zero of the 
equation above are unphysical as $Q<0$.
Therefore, an antibrane moving in this background will never have
enough energy to escape the ``gravitational attraction" due to the
geometry, and will stay in a bound state, bouncing back
and forth between the radii $\bar\eta_- <\eta <\bar\eta_+$, without
ever reaching the singularity.
In figure \ref{QhubbleKTq-1}$a$, the zero momentum case is shown 
in dashed lines.

\item $l\ne0$. The qualitative behaviour of the anti-D3-brane
trajectories when the angular momentum is turned on is the same as
in the zero  momentum case. 
This can be easily understood since  the number of
zeros of $Q$ stays the same, although the location 
changes, according to the equation
\be
\frac{c}{\eta^4}\,(1+b\,\ln{\eta}) - \frac{l^2}{\eta^2} = 2 \,.
\ee 
In figure \ref{QhubbleKTq-1}$a$, we show how the position of these
two zeros changes as we increase the value of the angular momentum. 
Therefore, the  trajectories of the antibrane are  as before:
An anti-brane wandering 
in the throat of KT, is always {\em constrained} to stay near the throat, 
oscillating between the two zeros of $Q$. 
As we will see in the next section, this behaviour will be 
reflected in a cyclic kind of expansion, when seen from the 
antibrane observer's point of view. 

\end{itemize}

\FIGURE[ht]{\epsfig{file=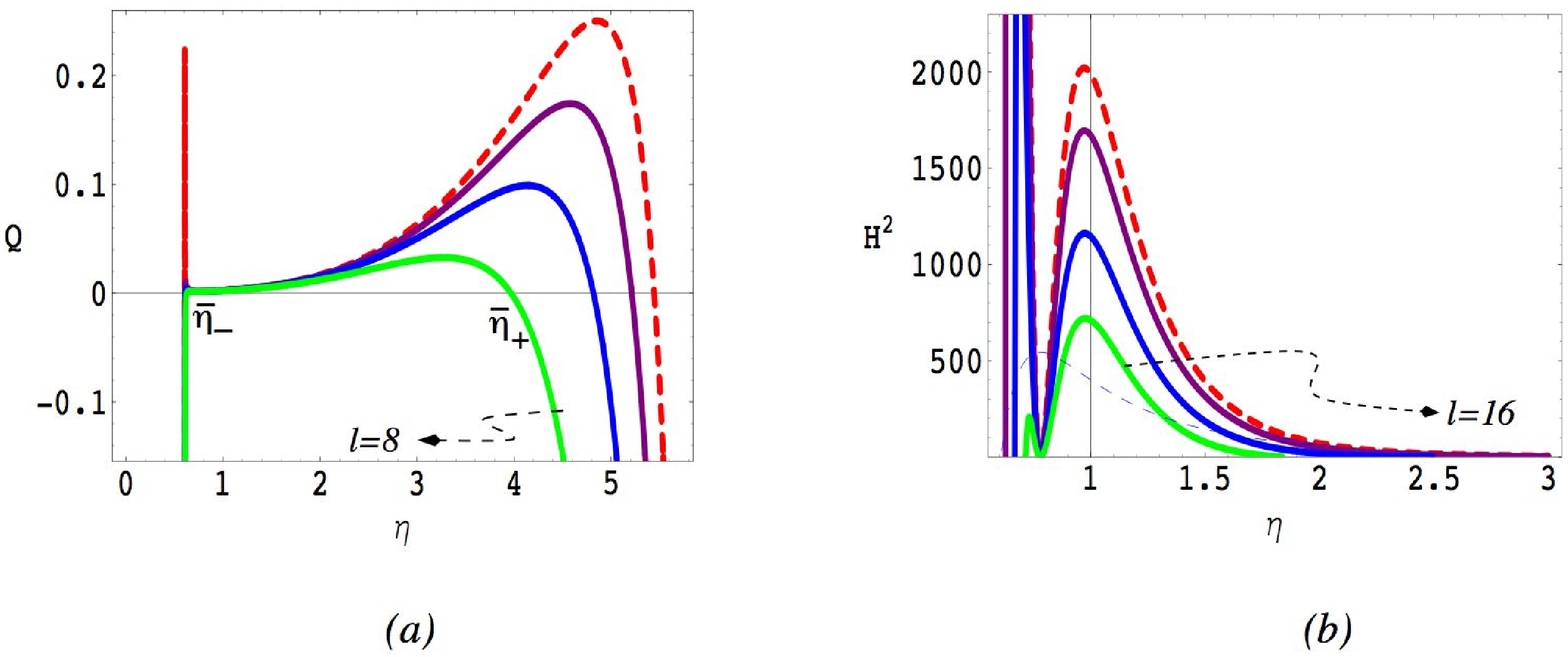, width=.9\textwidth}%
\caption{Variation of $Q$, $(a)$, and $H^2$, $(b)$, for the case of an 
anti-brane moving in the KT background. In this case, even for $l=0$, 
the quantity $Q$ has two roots, $\bar\eta_\pm$. When the angular
momentum is added, the physical region where $Q>0$ diminishes,
till it eventually disappears. Notice that for $\eta>\eta_+$, 
where $\eta_+$ is the biggest root of $Q$, there is no physical 
region. In terms of the Hubble induced expansion, these two 
roots give {\em only cyclic} universes for an antibrane.  
\label{QhubbleKTq-1}}}


\subsection{Induced Expansion in KT: Cyclic/Bouncing Universe}

The induced expansion that a brane (antibrane) observer experiences 
as they move in the KT background provides  novel examples of bouncing
and, more interestingly, cyclic brane universes, as we show below. 
For the KT background, eq.~(\ref{hubble1}) becomes
\be\label{friedkt}
H_{ind}^2= \left(\frac{h'}{4\,h^{3/4}} \right)^2 \left[ 
\varepsilon\,(h\,\varepsilon +2q) 
-\frac{l^2}{\eta^2} \right] = \left(\frac{h'}{4\,h^{3/4}} 
\right)^2 ( h\,\varepsilon + q)^2 \, Q
\,.
\ee
As we already mentioned, the angular momentum provides a negative
contribution that can be compared to  a positive curvature  
contribution in the standard FRW cosmology, thus, one can expect 
to get bounces once it is nonzero. In fact, 
the behaviour is far more interesting. 
In what follows, we discuss the generic behaviour of a 
brane/anti-brane moving in this background and illustrate it 
with several pictures.

\FIGURE
{\epsfig{file=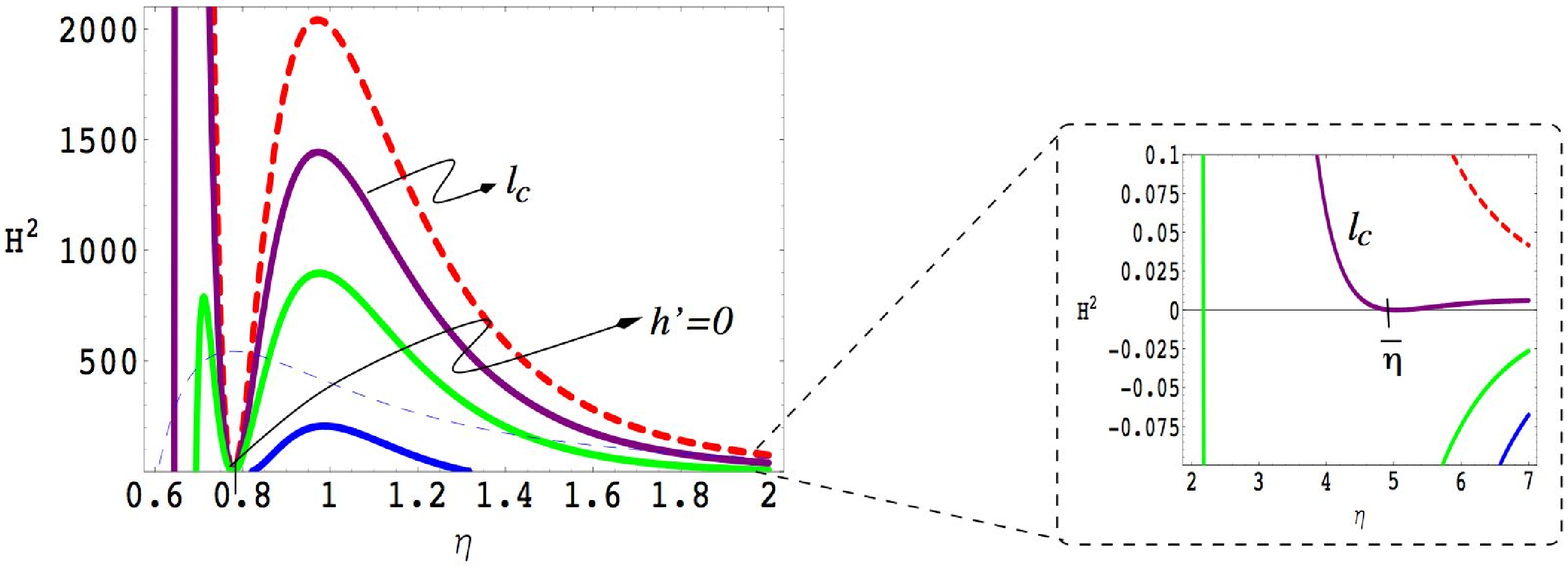, width=.9\textwidth}
\caption{Plot of $H^2(\eta)$ given by (\ref{friedkt}) for a brane. 
In the plot we take  the same values for the parameters as in 
figure \ref{QKT1}. The dashed curve corresponds to zero angular 
momentum, whereas the purple corresponds, as indicated, to $l_c$. 
The value of $\bar\eta$, where $Q=0$ for this curve (see figure \ref{QKT1}), 
is shown by zooming in on the curves around that point. 
The point at which $h'=0$ is $\eta'\sim 0.8$. The green curve 
corresponds to $l=19$. For this value of the angular momentum, the 
three first zeros of $H^2$ are clearly seen. The fourth lies 
outside the region shown in the plot. 
} \label{hubbleKT}}

\bigskip
\ni
{\it  Brane Expansion ($q=1$)}
\bigskip

\noindent
We now analyze the case of a moving brane. The type of induced
expansion that a brane observer will experience  is 
inferred from our previous discussion on the brane dynamics. We
describe each case in detail and again we illustrate our results
graphically with a numerical example. 
For the brane, the behaviour is graphically depicted in 
figure \ref{hubbleKT}.

\begin{itemize}

\item $l=0$.  As we  saw in the previous section, the quantity $Q$ is always
positive and has no zeros. However, it is still possible for the 
cosmological expansion to experience a
bounce if $h'=0$. 
This happens at only one value of $\eta$, $\eta'$, where $h$ has 
a maximum (see fig.~\ref{hKT}).  Therefore, a brane observer 
actually experiences a contracting Universe, as the brane falls 
towards $\eta'$, which then bounces and re-expands, and in fact
hyperinflates toward a final cosmological singularity in finite time.

\item $l\ne 0$. In this case, $Q$ can have up to three  
different roots, depending on 
the values of $l$. Since $h'$ has a zero at $\eta'$, the induced
Hubble parameter (\ref{friedkt}) can have a maximum of four zeros,
giving rise to a  rich structure. We consider each case separately. 

\begin{itemize}

\item $l_{min}<l<l_c$. In this case, $H^2$ has  two zeros. 
One is located at $\eta'$ and the other at $\bar\eta_1$
(one can check that $\bar\eta_1 <\eta'$). This is also clear from the plot 
of $H^2$ in figure \ref{hubbleKT}.  Therefore, a brane coming 
down the throat, will experience a contracting phase moving to an expanding
phase at $\eta'$ and then enter another contracting phase 
at $\bar \eta_1$, expanding again at $\eta'$.
 
\item $l=l_c$. As we increase the angular momentum to the 
critical value where $Q$ has two roots, a new zero for $H^2$ 
arises. Thus there are three turning points where 
$H^2=0$. In figure \ref{hubbleKT}, we zoom into the region where 
the zero of $Q$ at $\bar\eta$ appears. As we have already remarked, 
these are two possible brane trajectories. The first corresponds to 
a contracting phase (as the brane moves in from the UV) asymptoting
a steady-state Universe orbiting at $\eta=\bar \eta$. The other 
corresponds to a bouncing Universe,
oscillating between contracting and expanding phases.

\item $l>l_c$. In this case, there are four turning  points, 
where $H^2=0$. This gives rise, inevitably, to cyclic universes. 
A brane coming from  values of $\eta>\bar\eta_+$, will reach 
the turning point and bounce back to the UV region corresponding 
to a contracting/expanding cosmology. However,  
a brane moving in the region $\bar\eta_1<\eta<\bar\eta_-$ will 
experience a {\em multi-cyclic expansion}. 
In figure \ref{hubbleKT15} we plot $H^2$ for $l=15$ where 
the typical behaviour of the brane evolution 
along the throat is clearer.

\end{itemize}
\end{itemize}
 
\FIGURE
{\epsfig{file=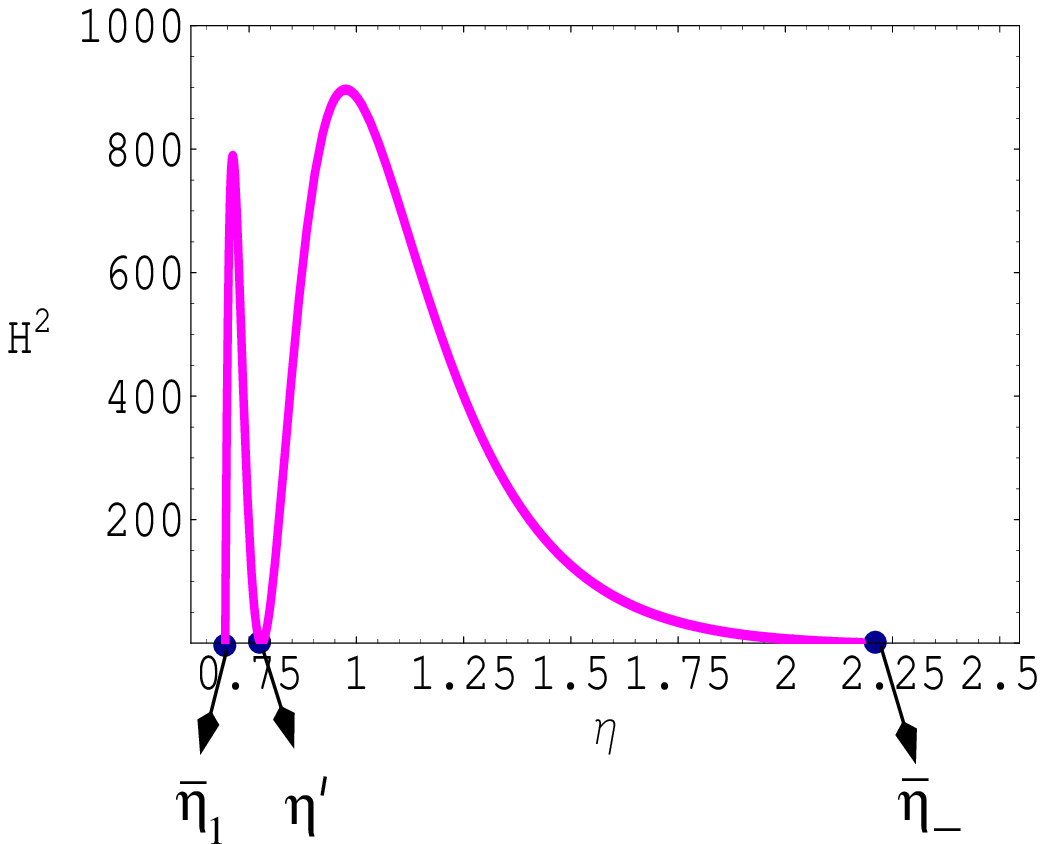, width=.6\textwidth}
\caption{Plot of $H^2(\eta)$ for $l=15$. The position of the three
first zeros of $H^2$ are shown. The two associated cyclic universe
regions are shown clearly.} \label{hubbleKT15}}

\bigskip
\ni
{\it Anti-Brane Expansion ($q=-1$) }
\bigskip

\noindent
The situation in the case of a wandering  anti-brane is very simple
to understand from the analysis of the trajectories discussed in the
last section. $Q$ has two roots, or turning points,
independent of the value of the angular momentum, which disappear
as $l$ becomes too large. Therefore, the induced expansion in this
case has the same qualitative form for $l=0 $ and $l\ne0$.
The form of the Hubble parameter  for the antibrane is shown in figure
\ref{QhubbleKTq-1}$b$.

As with the brane, there is another solution to $H^2=0$ coming from $h'=0$.
Therefore, $H^2$ has a total of 3 zeros, or
turning points, as can be seen from figure
\ref{QhubbleKTq-1}$b$. Since $H^2$ becomes negative after we
reach the (biggest) solution of $h=2$ ($Q=0$) at $\bar\eta_+$,
then, an antibrane observer will always experience a 
cyclic expansion, moving, as in
the case of the brane, between the two solutions of $Q=0$, without
touching  the singularity.


\section{Klebanov-Strassler Background}

The last  background we study in detail  is given by the 
Klebanov-Strassler geometry, a regularised version of the KT throat
studied in the previous section based on the deformed conifold. 
The metric ansatz for this background
is slightly more complicated than in the previous case, allowing for 
differential warping of the angular coordinates. The brane trajectories
are also rich, exhibiting bouncing and cyclic 
behaviour,  depending on the brane energy and angular momenta.

\subsection{The solution}

We now consider  the  regularisation of the KT  solution due to 
Klebanov and Strassler (KS) \cite{KS}.
It  describes the geometry due to a configuration of (a large number) 
$N$ D3-branes and (a large number) $M$ wrapped D5-branes. A cartoon 
of the regularisation of the KT solution is shown in figure \ref{KTKS}.

\FIGURE[ht]{\epsfig{file=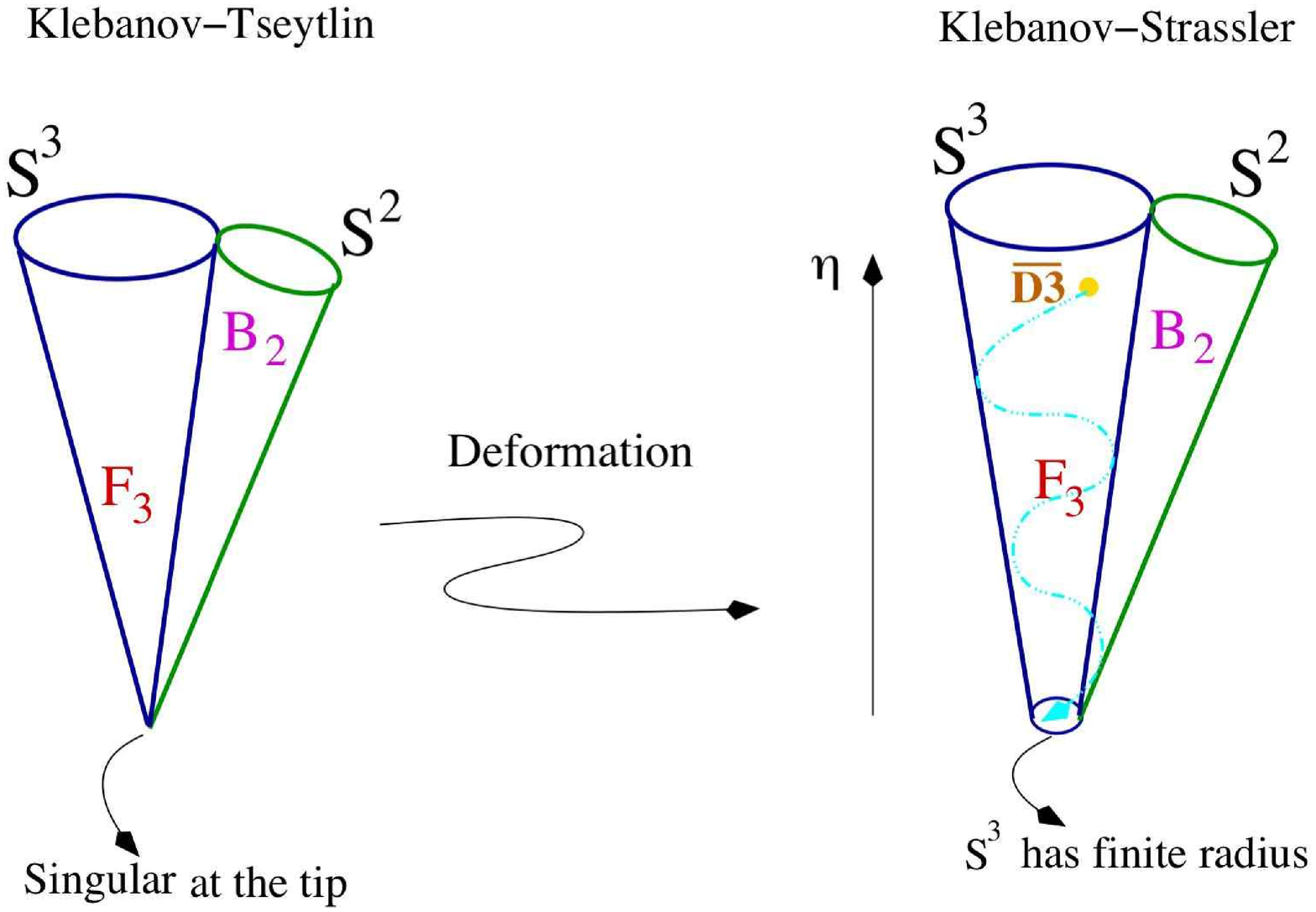, width=.6\textwidth}%
\caption{Cartoon representation of the KS resolution of the 
KT singular background.  \label{KTKS}}}

The metric and the other fields are given by 
\be\label{KS1}
ds^2_{10} = h^{-1/2}(\eta)\,dx_\mu dx^\mu +h^{1/2}(\eta) ds^2_6\,,
\ee
where the 6D metric is now replaced by that
of the {\em deformed conifold} (see ref.~\cite{mina})
\bea\label{KS2}
ds^2_6 &=& \frac{\epsilon^{4/3}}{2}\,K(\eta)
\Bigg[\frac{1}{3\,K(\eta)^3}\,\left\{d\eta^2 
+ (g^5)^2\right\}  \nn \\ 
&& \hskip1cm   + \cosh^2{(\eta/2)} \left\{(g^3)^2 +(g^4)^2\right\}
+\sinh^2{(\eta/2)} \left\{(g^1)^2 +(g^2)^2 \right\}   \Bigg]\,,
\eea
where the $\{g^i\}$ basis is defined as before and  $\epsilon$ 
is a constant that measures the deformation of the conifold. The 
other fields are  given by
\bea
&& B_2 = \frac{g_s\,M\,\alpha'}{2} \,\left[  f(\eta)\,g^1\wedge g^2 
+ k(\eta)\, g^3\wedge g^4  \right]\,, \nn\\
&& H_3= \frac{g_s\,M\,\alpha'}{2} \,\left[ d\eta\wedge (  f'(\eta)\,g^1
\wedge g^2 
+  k'(\eta)\, g^3\wedge g^4 )
+ \frac{1}{2}\left( k(\eta) - f(\eta) \right) g^5\wedge \left(  
g^1\wedge g^3
+ g^2 \wedge g^4  \right)\right]\,, \nn \\ 
&& F_3= \frac{M\,\alpha'}{2} \,\left[ \left( 1- F(\eta)\right)
g^5\wedge g^3  \wedge g^4 + F(\eta)\, g^5\wedge g^1\wedge g^2
+  F'(\eta)\, d\eta\wedge \left(  g^1\wedge g^3
+ g^2 \wedge g^4  \right)\right]\,,\nn \\ 
&& \tilde F_5 = {\mathcal F}_5 + \star {\mathcal F}_5 = 
B_2\wedge F_3 + dC_4 \nn \\
&& dC_4 = g_s^{-1}
d(h^{-1})\wedge dx^0\wedge dx^1\wedge dx^2\wedge dx^3\nn\,.
\eea
Again, the dilaton and the RR zero form vanish in this solution. 
The explicit form of the functions appearing above is:
\bea
&& F(\eta) = \frac{\sinh{\eta}-\eta}{2\sinh{\eta}} \,,\\
&& f(\eta) = \frac{\eta\,\coth{\eta}-1}{2\sinh{\eta}}(\cosh{\eta}-1)\,,\\
&& k(\eta) = \frac{\eta\,\coth{\eta}-1}{2\sinh{\eta}}(\cosh{\eta}+1) \,, \\
&& K(\eta) = \frac{\left( \sinh{(2\,\eta) - 2\,\eta}\right)^{1/3}}{2^{1/3}
\,\sinh{\eta}} \,,\\
&&  h(\eta) = (g_s M  \alpha')^2  2^{2/3} \,\epsilon^{-8/3}
\int_\eta^\infty{dx\,\frac{x\,\coth{x} -1 }{\sinh^2{x}}\,
\left( \sinh{(2\,x) - 2\,x}\right)^{1/3} }\,.\label{hks}
\eea  
Although the integral above, conventionally denoted as $I(\eta)$, 
cannot be computed analytically, one can readily work out the two  
important limits of the solution as $\eta\to0$ and $\eta\to \infty$, 
\cite{KS,coneuse}.  One can 
check that near the bottom of the throat (IR), the internal metric takes the
form of a 3-sphere with finite radius, plus a 2-sphere which shrinks to zero
size. At the other end of the throat (UV), the metric takes the
form of the KT solution, asymptoting AdS spacetime. 
The relevant limits of the integral above are
\be
I(\eta\to 0) \to a_0 + {\mathcal O}(\eta^2) \,\,; \qquad 
I(\eta\to \infty) \to 3\cdot2^{-1/3}(\eta - 1/4)\,e^{-4\eta/3}\,,
\ee 
where $a_0 = 0.71805$ \cite{coneuse}.


\subsection{Brane Evolution and Physical Consequences}

We are  again dealing with a the Lagrangian of the form (\ref{L})
and we take  the brane to be moving along the same cycle as in KT. 
Therefore the speed in the present case becomes ($\theta_1=\theta$):
\be
v^2 = A(\eta)\,\left[ \dot \eta^2   + \dot \psi^2 \right]
+B(\eta) \,\dot\theta^2
\ee
where 
\be
A(\eta)= \frac{\epsilon^{4/3}}{6K^2}\,, \qquad \quad 
B(\eta) = \frac{\epsilon^{4/3} K}{4}\,\left[\cosh^2{(\eta/2)} + 
\sinh^2{(\eta/2)}\right]
\ee
The two conserved angular momenta along $\psi$ and $\theta$  are given by 
\bea
l_{\psi} = \frac{A\,\dot \psi}{\sqrt{1-h \,v^2}} \,; \qquad \qquad
l_{\theta} = \frac{B\,\dot \theta}{\sqrt{1-h \,v^2}}  \,,
\eea
and the canonical momentum associated to $\eta$ is 
\be
\rho_\eta = \frac{A\,\dot \eta}{\sqrt{1-h \,v^2}}  
\,.
\ee 
Using  this information, the energy (\ref{energy2}), becomes
\be\label{energyks}
\varepsilon = \frac{1}{h} \left[ \sqrt{\frac{1+ h\,\ell^2(\eta)}
{1-h\,g_{\eta\eta}\,\dot\eta^2}} - q \right]
\ee
where 
\be
\ell^2(\eta)= 
\frac{l_\psi^2}{A} +\frac{l_{\theta}^2}{B} \,.
\ee
The time 
evolution of the brane along the radial coordinate $\eta$ (\ref{etadot1})
is then determined by
\be\label{etadotKS}
\dot\eta^2 = \frac{\left[  \varepsilon(h\,\varepsilon + 2q) 
- \ell^2(\eta)\right]}
{A (h\,\varepsilon+q)^2}= Q
\,.
\ee

We now study the behaviour of the brane (or anti-brane), as it 
moves in this background. We find that, although the motion 
has many similarities with the KT case, there are some important 
differences. 

\bigskip
\ni
{\it Dynamics of a brane ($q=1$)}
\bigskip

\noindent
In order to study the form of the brane trajectories in the KS
background, we numerically evaluate the metric function (\ref{hks})
for consistent values of the parameters. We take  
$\alpha^\prime=1$ and $g_s M$, $g_s N \sim \mathcal O(10)$, 
and plot the relevant metric functions in figure \ref{hKS}. 
Moreover, note the presence of a $g_{\eta\eta}=A(\eta)$, and
the effect of the differential warping $A(\eta)$ and $B(\eta)$. 
Although these functions become proportional
at large $\eta$ ($B \propto 3 A/4$), their precise form at finite
values of $\eta$, will be important when we turn on the angular
momenta $l_i$.  Notice that, in contrast with the KT case, 
$h>0$ for all values of $\eta$, and tends to zero for large values of
$\eta$ only. This will have important consequences later. 
Let us  look in more detail at the brane evolution in the KS
background. We concentrate on the case $\varepsilon=1$, and show
the results for $Q$ as we change the angular momenta in figure \ref{QKS}.

\begin{itemize}
 
\item  $l_i =0 $. When both momenta are zero, $Q$ becomes
$$Q=\frac{h+2}{A(h+1)^2}\,.$$ Therefore, at $\eta=0$, it has a
constant positive value,  $Q_0=Q(0)$. Since
$h\,,A>0$, in this case, $Q>0$ for all $\eta$ values. 
Therefore a D3-brane is free
to move along the entire geometry without restrictions. It will
do so by increasing its radial velocity as it moves toward the tip of the
throat, but  decreasing it  as it comes closer to it (see
fig.~\ref{QKS}). 

\item   $l_i\ne 0$. As we turn on the angular momentum, the
brane's radial speed decreases relative to no angular contribution,
eventually reaching a zero value for large enough angular
momenta. The turning points where this can happen, are given by
the equation 
$$h-\ell^2(\eta)=-2 \,.$$
Just as in the case of the KT  background, one can check that the
maximum number of solutions for this equation is three. In particular,
it is the contribution of $l_\theta$ that provides a third
solution\footnote{ That is, if $l_\theta=0$, $l_\psi\ne0$ 
only two solutions are possible. However, if $l_\psi=0$, and
$l_\theta\ne0$,  a maximum of three zeros are possible.}.  
This can be seen from figure \ref{QKS}.
Therefore, without loss of generality, we can  extract the most
general type of trajectories by  concentrating  on the case
$l_\psi=0$, $l_\theta\ne0$.
The behaviour is  very similar to the KT one, with some interesting 
differences, which we now describe.

\begin{itemize}
 
\item $l_\theta<l_c$. In this case, there are no solutions
to $Q=0$, thus, the trajectories are qualitatively the same as in
the zero momentum case, the only difference being that 
the  radial speed  decreases via a centrifugal repulsion.

\item  $l_\theta = l_c$. At this point, the equation $Q=0$ has
a solution at $\bar\eta$, where the radial speed and acceleration
vanishes.  A brane coming from the UV region, will asymptote
a circular orbit at $\bar\eta$. 
 
\item $l_\theta >l_c$. When the angular momentum is increased
above the critical value, three turning points can appear at
$\bar\eta_1$, $\bar\eta_\pm$, as shown in
figure \ref{QKS}$b$. Therefore, there are two separate regions
where the brane can move. For values of $\eta>\bar\eta_+$ the brane
coming  down the throat, will reach a 
maximum distance from the tip, where its radial speed is zero, 
bouncing back to the region of large $\eta$ values. Moreover, a
new internal region appears between, $\bar\eta_1<\eta<\bar\eta_-$, where
the brane motion is bounded, as shown in figure \ref{QKS}.

\end{itemize}

\end{itemize}

\FIGURE[ht]{\epsfig{file=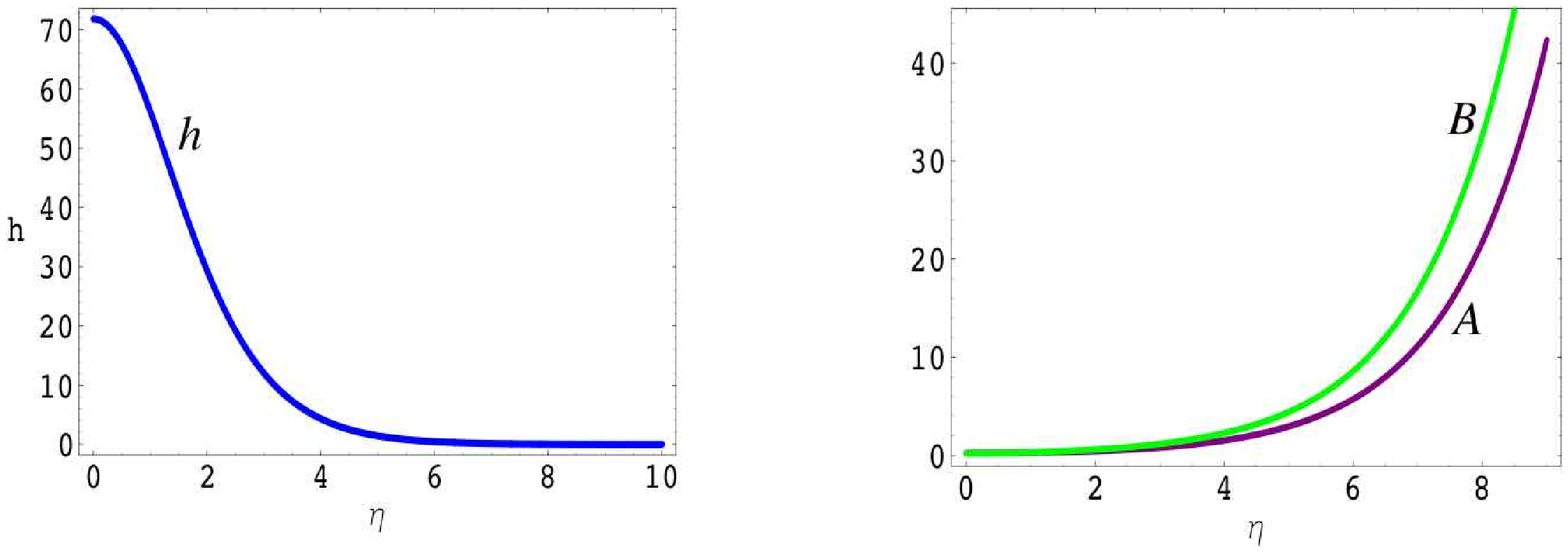, width=.6\textwidth}
\caption{Plot of $h(\eta)$ given by Eq.~(\ref{hks}), and functions $A$
and $B$.  In the plot we take $\alpha^\prime=1$ and $g_s M$, $g_s N
\sim \mathcal O(10)$. }\label{hKS}}

\FIGURE[ht]{\epsfig{file=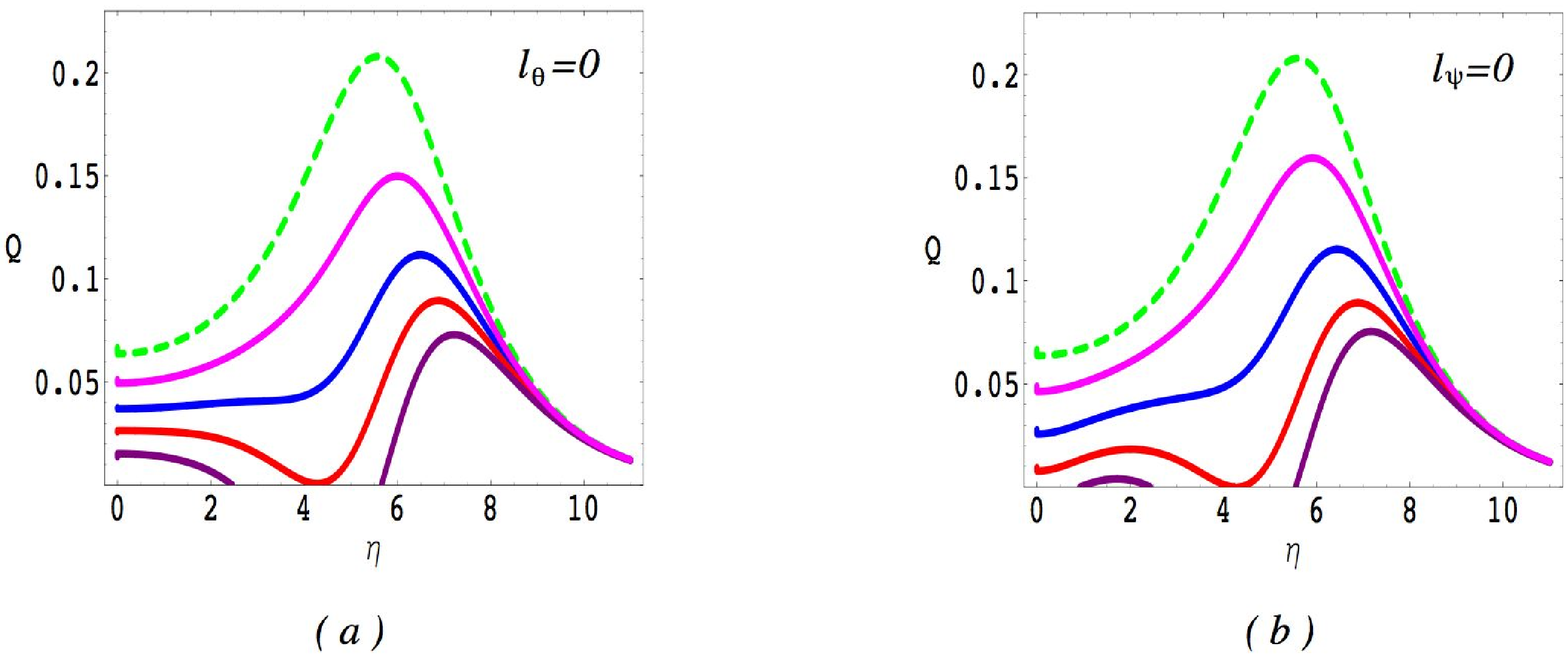, width=.9\textwidth}
\caption{The function $Q$ for a brane moving in the KS background as
the angular momenta changes. In
picture $(a)$  we  keep $l_\theta=0$, and allow $l_\psi$ to
vary, whereas in $(b)$ we keep $l_\psi=0$ and change $l_\theta$. 
It is clear that in the second case, a second zero can appear near the tip
of the deformed conifold. We took the same values for the parameters
as in figure \ref{hKS} and $\varepsilon=1$.}\label{QKS}}

\FIGURE[ht]{\epsfig{file=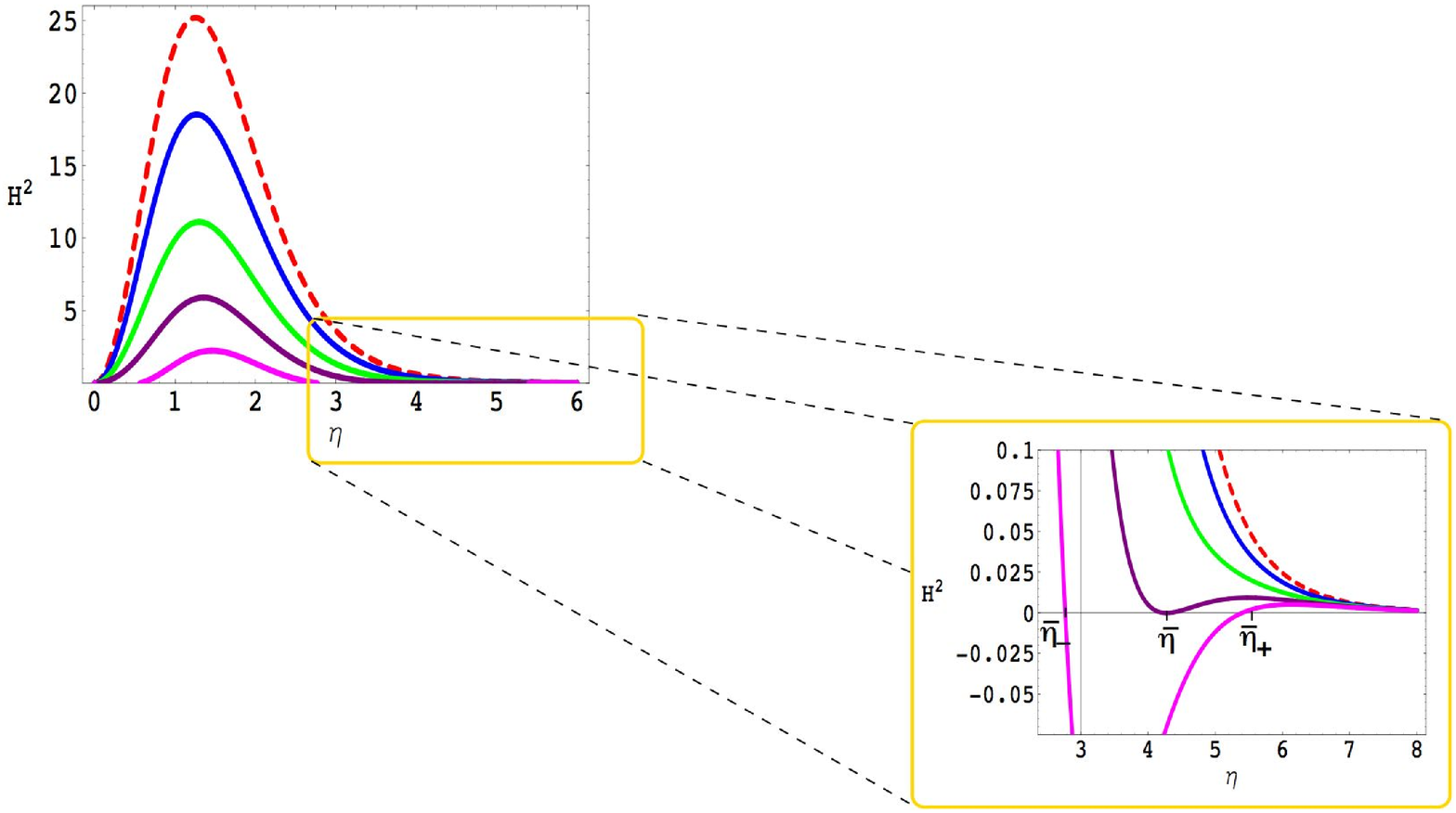, width=.9\textwidth}
\caption{Induced expansion as seen by a brane observer when the brane moves
along the KS throat. The parameters are chosen as in
figure \ref{hKT}. There is always a zero of $H$ at
the origin of the regular KS geometry. We zoom in on the right
region where it is clear that the zeros of $Q$ are located, as we
increase the angular momentum. As explained in the text, for
$l_\theta>l_c$ there are no zeros for $H$ apart from the origin. 
For $l_\theta=l_c$ a new zero arises, at $\bar\eta$. This zero splits
into $\bar\eta_\pm$ as we increase $l$ above its  critical
value. For the values of the parameters we are taking on the plot,
$l_{c} \sim 3.764$. }\label{hubbleKS}}


\bigskip
\ni
{\it Dynamics of  an Anti-Brane ($q=1$)}
\bigskip

\noindent
The case of an antibrane wandering in the KS throat is very similar
to the case of the KT background. The trajectories of the 
antibrane behave very similarly, independent of the value of the angular
momentum. There is however, a small 
range of parameters, where a small variation can arise. Let us now
study the trajectories in detail. We concentrate, as 
in the previous case, only on 
the case where $l_\psi=0$, since, 
as we explained above, it is $l_\theta$ which
provides the rich structure of the trajectories. 

\begin{itemize}
 
\item  $l_i =0 $.  When the angular momenta vanishes, there is  a 
single solution to $Q=0$, that is, $h=2$, at $\bar\eta$, say,
as can be seen by the form of $h$  (fig.~\ref{hKS}).
Therefore, the antibrane is constrained  to move in the region between the
resolved tip of the conifold, and $\bar\eta$.
This case is represented in figure \ref{QhubbleKS}$a$ by the dashed line. 
One can compare this to the KT case in figure \ref{QhubbleKTq-1}$a$.

\item   $l_\theta\ne 0$. When the angular momentum is turned on,
the equation $Q=0$ implies\footnote{Again, adding the other angular
momentum, $l_\psi$ gives the same results and is trivially included.} 
$$h-\frac{l_\theta^2}{B}=2\,. $$
Given the form of $h$ and $B$, one can easily check that, 
for generic values of the angular momentum, 
there is always one solution, just as in the previous case. 
However, the position of the zero 
of the radial speed shifts to smaller values of $\eta$, restricting
the region of solutions for the antibrane trajectories. Also,
the value of the radial speed decreases (see fig.~\ref{QhubbleKS}$a$). 
As we increase the value of the angular momentum, the radial speed
continues to  decrease, until, 
the region where the antibrane can move, vanishes. 
As this happens, there is a small range of values of $l_\theta$
where a bounded state with small radial speed  arises. 
This happens when the solution to the 
equation above has  two solutions $\bar\eta_\pm$. 
This region will become clearer when we look at the 
induced expansion in the next subsection. 
In figure \ref{QhubbleKS}$a$ we show how
the antibrane radial speed changes with the angular momentum ($l_\theta$).

\end{itemize}


\subsection{Induced Expansion in KS: Cyclic/Bouncing Universe}

The induced expansion that  a brane (antibrane) observer experiences 
in the present case, can be already intuited 
using our experience with the AdS  and the KT backgrounds.  
Using (\ref{etadot1}) and (\ref{hubble1})  in the
KS case,  we find  the effective 4D Friedmann equation:
\be\label{friedks}
H^2_{ind}= \left(\frac{h'}{4\,h^{3/4}} \right)^2 \left[ 
\frac{\varepsilon\,(h\,\varepsilon +2q)}{A} 
-\frac{l_{\psi}^2}{A^2} 
-   \frac{l_{\theta}^2}{A\,B} \right] = 
\left(\frac{h'}{4\,h^{3/4}} \right)^2  (\varepsilon \,h +q)^2 \,Q
\ee
Therefore, just as in the previous two examples (AdS and KT) 
the angular momentum provides an 
important negative contribution to the Friedmann effective 
equation which again gives rise to
bouncing and cyclic universes. 
We now consider these solutions in detail and, as in the previous
analysis, we compute 
numerically a concrete example, which illustrates our results. 

\bigskip
\ni
{\it Moving Brane}
\bigskip

\noindent
We start with the case of a brane 
($q=1$). The generic behaviour is very similar 
to the KT case. The main difference is that there is no
singularity in the background,
and the form of $h$, and thus $h'$ changes. We show this case in figure 
\ref{hubbleKS}. 

\begin{itemize}

\item $l_i=0$. The vanishing momentum case is shown with a  
dashed line in figure \ref{hubbleKT}.
Following our general analysis in section \ref{inducexp}, we 
know that in order to
get  bounces, there should be solutions to $H^2_{ind}=0$.
For vanishing angular momentum,
these solutions are provided by $h'=0$. In the KS case, this has a 
single solution at $\eta=0$.
Therefore, there is a single zero for the induced Friedmann equation.
Physically, a brane
coming from the UV region of the KS background, arrives at the end of
the throat at $\eta=0$, where it passes through, returning to the UV region 
along an antipodal direction.
This solution was studied in \cite{km}.

\item $l_i\ne 0$. When the angular momentum is turned on, there is a maximum 
of three zeros for $Q=0$ at $\eta\ne0$. Besides these, there is the 
extra zero of $h'$ at $\eta=0$. 
Thus, (\ref{friedks}) has a maximum of four zeros, which represent
bouncing points for the induced cosmology. 
As can be seen from figure \ref{hubbleKS}, there are three
different situations, depending on the amount of angular momentum,
with similar behaviour to the KT case.

\begin{itemize}

\item  $l_\theta < l_c$. For these values of the angular
momentum, the situation is the same as for the case of no angular
momentum, that is the brane experiences a single bounce at the
origin, with  $H^2$ reduced relative to $l=0$.  

\item  $l_\theta = l_c$. When $Q$ has a single zero at
$\bar\eta$, the universe once more asymptotes a steady state cosmology
in orbit at $\eta=\bar\eta$.

\item  $l_\theta > l_c$. For larger values of  $l_i$, there
are four solutions to $H^2=0$: $0$, $\bar\eta_1$, $\bar\eta_\pm$. 
Then, two different regions for the brane expansion appear. 
The brane observer can either experience a bounce at the largest zero
of $Q=0$, $\eta=\bar\eta_+$, or
it can bounce back and forth, between the two smaller zeros of $Q=0$,  
$\bar\eta_{1}\le\eta\le\bar\eta_-$. Therefore a brane observer will
experience a cyclic expansion.      
(Note that the region between $0<\eta<\bar\eta_{1}$ is not physical.)

\end{itemize}
\end{itemize}

\bigskip
\ni
{\it Moving Anti-Brane }
\bigskip

\noindent
The case of an antibrane ($q=-1$) 
moving in the KS background may be anticipated 
from our previous analysis of AdS and KT backgrounds. Indeed the physical 
behaviour changes only slightly. We illustrate our results in figure
\ref{QhubbleKS}$b$. One can compare this picture with that of the
KT background,  figure~\ref{QhubbleKTq-1}.

\begin{itemize}
 
\item $l_i=0$. For vanishing $l_i$, we saw that $Q=0$ has
always a single solution.  
Therefore, $H^2=0$ has two solutions. Physically, an antibrane will
always experience a cyclic expansion, bouncing back and forth between
zero (where $h'=0$) and the single root of $Q$ at $\bar\eta$.

\item  $l_\theta \ne 0$. As we turn on the angular momentum, in
general, the trajectories have the same qualitative  behaviour as
before. That is, the antibrane
experiences a cyclic expansion. The effect of the angular 
momentum is to decrease the value of the Hubble parameter induced on
the brane, until the  region where $H^2>0$ vanishes. 
Before this happens, there is a small region of values of the angular
momentum, 
where two roots of $Q$ appear at $\bar\eta_\pm$, as we saw before. In
those cases, the cyclic brane universe  
exists between these two roots of $Q$, and the region between $\eta=0$
and the smaller root of $Q$ is not physical.

\end{itemize}

\noindent
As it is clear from our discussion, cyclic behaviour is quite 
generic, once we turn on angular momentum.


\FIGURE[ht]{\epsfig{file=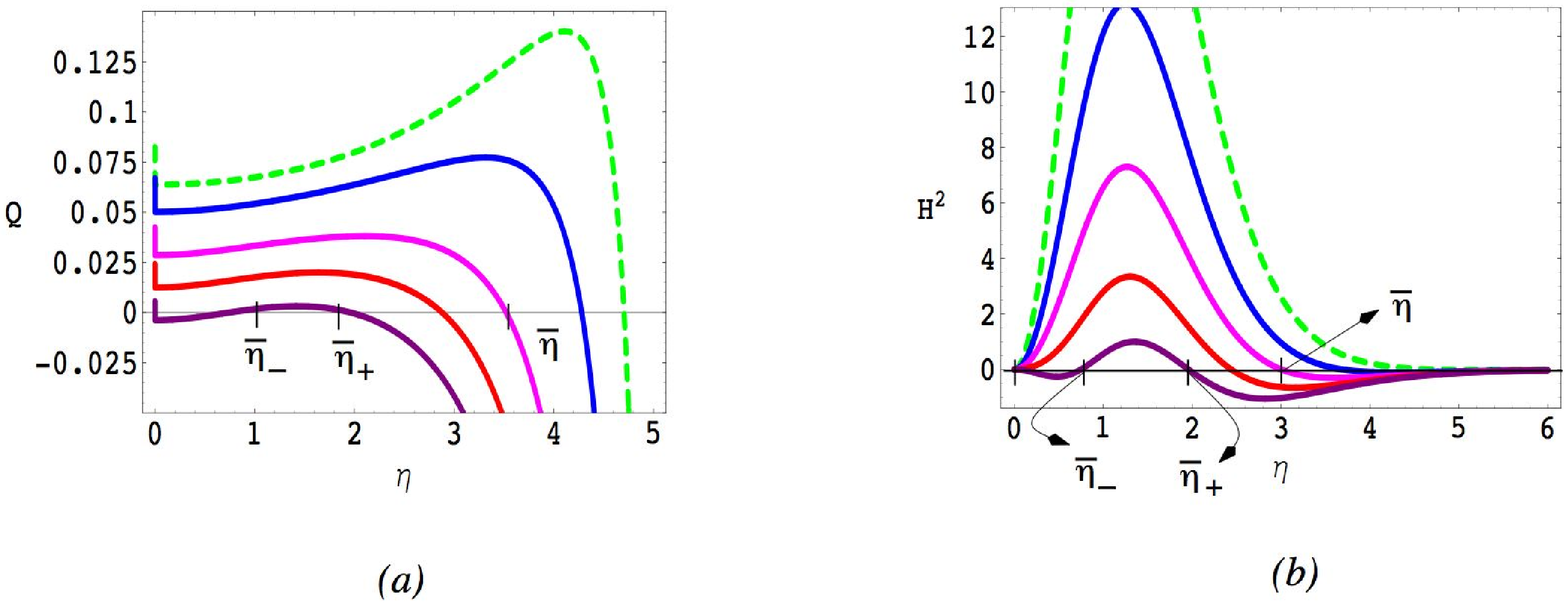, width=.9\textwidth}
\caption{The anti-brane kinetic function $Q$ ($a$) 
and Hubble induced expansion $H^2$ $(b)$, for the same values  
of parameters as in figs.~\ref{hKS}--\ref{hubbleKS}, as the angular
momentum $l_\theta$ 
changes. In both pictures, the dashed line corresponds to 
vanishing angular momentum. The value of the angular momentum for
which a new zero of $H$ arises, for the values we are taking is
$l_\theta\sim 4.02$.}\label{QhubbleKS}}

\section{Discussion}

In this paper, we studied in detail the dynamics of probe D3-branes and 
antibranes, as they move along the transverse directions of a warped 
supergravity background with fluxes. In particular, 
we allowed the brane (antibrane) to move in more than one of 
the transverse internal coordinates: that is, along
angular as well as radial coordinates. In so doing, we have uncovered 
interesting novel trajectories that a brane/antibrane can take in these
backgrounds.

We began by considering the pure $AdS_5\times S_5$ geometry as a 
classic and well known example. This has been analysed in
the literature for the case of a moving brane, and we have completed the 
picture by including the antibrane. We then modified the
background with the addition of fluxes, considering respectively the 
Klebanov-Tseytlin and Klebanov-Strassler backgrounds.
Not surprisingly, the dynamics of the brane on these three backgrounds 
presents some similarities, due to the fact that the warping has similar 
features in all the three examples.  However, a more careful study revealed 
interesting differences. These differences are related both to the varying
forms of the warp factors in the three cases, and to the different 
manifolds in which the angular coordinates are compactified. For a probe 
brane in the three examples, we found that the type of trajectories 
depend crucially on the value of the angular momentum.

It is possible to distinguish between three qualitatively different 
possibilities for the brane dynamics. In the first case, the brane moves 
freely through the whole spacetime, eventually reaching the tip of the throat 
(either the AdS horizon, the singularity or the regularised tip). In the 
second case, a brane coming from the UV experiences a bounce that brings it 
back and does not allow it to approach the tip of the throat. This case is
interesting because it shows that a brane that moves through singular 
backgrounds, like the KT, can have regular trajectories that avoid the 
singular parts of the geometry. In the last case, the brane is 
confined to move inside a bounded region, providing a novel example of cyclic 
motion inside the throat.
For a probe antibrane, the trajectories are qualitatively
independent of the value of the angular momentum
in the three examples that we considered. In the KT and KS
geometries, a turning point occurs before the tip of the throat.

We also studied the brane evolution as seen from the point of view of
a brane observer, or mirage cosmology \cite{kiritsis}.  
The trajectories found suggested the possibility of both bouncing and cyclic 
universes. Indeed, for the pure AdS throat, as was shown in
\cite{kiritsis,germani} for branes, bouncing universes arise naturally as 
we turn on the angular momentum, whereas they are always present for the
antibrane.

More interesting are the KT and KS backgrounds where a bouncing universe 
is {\em always} present. This, as we pointed out in the text, is due to 
the form of the warp factor. When we turn on the angular momentum, 
as well as bouncing, cyclic cosmologies also arise. Although we
 have studied  three specific backgrounds, we expect the same 
qualitative behaviour in a generic warped compactification.  
One of the nice features about having cyclic trajectories is that, whether 
or not they correspond in the final analysis to cyclic universes (which
depends on the effect of backreaction),
the brane experiences successive periods of acceleration during this motion,
which can feed into successive inflationary eras \cite{st}. 

Our analysis can be extended in several directions, for example, we have not
taken into account any mechanism for the stabilization of the 
K\"ahler moduli, and the consequences of this for the action of
the probe brane. This is an important issue for any genuine 
cosmological brane scenario in which we would require stabilization 
of all moduli. Initial work, \cite{malda}, indicates that the effect of
stabilization is to cause an additional attractive force towards the 
tip of the KS throat. Presumably, this will lift the critical values of
angular momentum for which there is sufficient centrifugal repulsion 
for a bounce in the throat to occur. We do not expect the qualitative
families of trajectories to be altered however. In particular, we would
still expect cyclic trajectories in the stabilized KS throat.

More importantly, we have neglected the gravitational
backreaction of the probe brane in our discussion of cosmology, in
other words, we have been taking the {\it mirage} point of view. Although
this is one possible way to proceed, it may not be   
the most satisfactory one. In the case of a simple codimension one
braneworld, such as Randall-Sundrum, the association between moving
branes and cosmology is proven. Because of the symmetry
of the geometry, and hence equations of motion, any isotropic perfect
fluid cosmological source can be added to the brane and fully supported
as a solution to the equations of motion via the Israel equations. However,
we already lose this concrete correspondence as soon
as we add one more dimension, even within the supergravity description
which, as we have discussed, is severely limited for the probe brane.
Thus, for codimension two and higher, there is no prescriptive way
to add energy momentum to a brane. 
Given this, it is difficult to 
get a concrete formulation of any mirage description as truly corresponding
to a cosmological universe containing real matter, and evolving
according to local physical principles. In fact, the `matter' in the mirage
scenario is simply that inferred from applying the standard Einstein-FRW
equations to the induced evolution, and is not the result of any
localised fields on the brane.
It is difficult to see how to address problems such as
reheating, particle production, or indeed whether we can
get a sensible description of the standard model on a probe brane.

For this reason, we have discussed the induced cosmology of our brane
trajectories in a somewhat limited fashion, as we do not believe that
computation of properties depending on a particle interpretation
(such as the spectral index of perturbations) will be robust to any
mechanism for embedding this interpretation into a more complete
gravitational description. As we have remarked, it appears that it
is not possible within the supergravity approximation to get a truly
gravitational description of a moving brane with matter in more than
one codimension (although, see \cite{cod2+} for progress in the case
of Gauss-Bonnet gravity), and given that one of the important novel
cosmological predictions of the Randall-Sundrum model was the addition
of $\rho^2$ terms to the Friedman equation, the lack of a rigorous
means to project gravity onto the brane is disappointing from the
point of view of a spacetime picture of string cosmology.
However, one can take a more pragmatic approach to backreaction,
and as a first step couple the DBI action we have 
studied to four dimensional gravity along the lines of \cite{Gary}, 
and analyse the resulting cosmologies as in \cite{st}. 
A full study of cosmology in this direction is currently underway.

Of particular cosmological relevance is whether the cycling trajectories
we have found in this paper will correspond to cyclic universes in the 
gravitationally coupled theory. In fact it is straightforward to show
that this is not the case. Computing $\rho$ and $p$ from the 
Lagrangian (\ref{L}), even allowing for a possible potential term
for the radial variable, demonstrates that the energy momentum obeys 
both the weak and null energy conditions, and hence a cyclic
evolution of the scale factor cannot arise. Note that this does not
mean that the brane cannot follow a cycling trajectory in the throat,
simply that this cycling trajectory cannot correspond to a cyclic
universe. The addition of the effective Einstein Hilbert term on the
brane breaks the direct correspondence between the scale factor of 
the four-dimensional cosmology and the radial position function of
the brane. Because of this, it is possible that a brane could have
an epoch of cycling in the throat,
before exiting into the asymptotic regime and standard big
bang cosmology.
It is interesting that this simple first step towards gravitational
back reaction gives such a radical modification of the mirage 
picture, however, it is always possible that higher order corrections
may negate this.

Finally, another interesting issue related to our results is the dual
field theory interpretation of the cycling trajectories:
this could provide further intriguing connections between cosmological
trajectories and their interpretation in the dual theory. 

\acknowledgments

We would like to thank Daniel Baumann, Cliff Burgess, 
David Mateos, Liam McAllister, Simon Ross, and Martin
Schvellinger for useful discussions. 
This research was supported in part by PPARC. DE, RG and GT 
are partially supported by the EU $6^{th}$ Framework Marie Curie 
Research and Training network ``UniverseNet" (MRTN-CT-2006-035863).
GT is also supported by the EC $6^{th}$ Framework Programme Research 
and Training Network MRTN-CT-2004-503369.
IZ is supported by a PPARC Postdoctoral Fellowship.

\end{document}